\documentclass[12pt, letter]{article}
\usepackage[margin=1in]{geometry}
\usepackage{setspace}
\usepackage[affil-it]{authblk} 

\RequirePackage{hyperref}
\usepackage{amsmath, amssymb,amsfonts}
\usepackage{amsthm}
\usepackage{subcaption}
\usepackage{bm, bbm,color}
\usepackage{graphicx} 
\usepackage{algorithm, algorithmic}
\usepackage{cleveref} 
\usepackage{natbib}
\usepackage{float}

\newcommand{\dbtilde}[1]{\tilde{\raisebox{0pt}[0.85\height]{$\tilde{#1}$}}}

\newcommand{\Rbb}{\mathbb{R}}

\newcommand{\Ccal}{\mathcal{C}}

\newcommand{\Ncal}{\mathcal{N}}

\newcommand{\norm}[1]{\left\|#1\right\|}

\newcommand{\INDSTATE}[1][1]{\STATE\hspace{#1\algorithmicindent}}

\begin{document}
\title{Bayesian Tensor-on-Tensor Regression with Efficient Computation}
\author{Kunbo Wang \quad Yanxun Xu* }
\affil{\footnotesize Department of Applied Mathematics and Statistics, Johns Hopkins University}
\date{}
\maketitle

\begin{abstract}
	We propose a Bayesian tensor-on-tensor regression approach to predict a multidimensional array (tensor) of arbitrary dimensions from another tensor of arbitrary dimensions, building upon the Tucker decomposition of the regression coefficient tensor. Traditional tensor regression methods making use of the Tucker decomposition either assume the dimension of the core tensor to be known or estimate it via cross-validation or some model selection criteria. However, no existing method can simultaneously  estimate the model dimension (the dimension of the core tensor) and other model parameters. To fill this gap, we develop an efficient Markov Chain Monte Carlo (MCMC) algorithm to estimate both the model dimension and parameters for posterior inference. Besides the MCMC sampler, we also develop an ultra-fast optimization-based computing algorithm wherein the maximum a posteriori estimators for parameters are computed, and the model dimension is optimized via a simulated annealing algorithm. The proposed Bayesian framework provides a natural way for uncertainty quantification. Through extensive simulation studies, we evaluate the proposed Bayesian tensor-on-tensor regression model and show its superior performance compared to alternative methods. We also demonstrate its practical effectiveness by applying it to two real-world datasets, including facial imaging data and 3D motion data. 
\end{abstract}



\section{Introduction}
Multi-dimensional arrays, also called tensors, are widely used to represent data with complex structures in different fields such as genomics, neuroscience, computer vision, and graph analysis. For example,  a multi-tissue experiment \citep{wang2019three} collects gene expression data in different tissues from different individuals, leading to three-dimensional arrays ($Genes\times Tissues \times Individuals$). Other notable examples include magnetic resonance imaging data (MRI, three-dimensional arrays),  functional MRI (fMRI) data (four-dimensional arrays), and facial images (four-dimensional arrays)\citep{vasilescu2002multilinear,hasan2011review,  guhaniyogi2021bayesian}. In this paper, we focus on the task of tensor-on-tensor regression that predicts one multi-dimensional tensor from another multi-dimensional tensor, e.g., predicting gene expression across multiple tissues for multiple individuals from their clinical/omics data with tensor structures.


One simple approach dealing with tensor-on-tensor regression is to turn   tensors into vectors, and then apply classic regression methods.
However, such a treatment introduces high-dimensional unstructured vectors and destroys the correlation structure of data, resulting in a huge number of parameters to be estimated and potentially significant loss of information. 
For example, to predict a response tensor of dimensions $N\times Q_1 \times Q_2$ from a predictor tensor of dimensions $N\times P_1 \times P_2$, the classic linear regression method requires estimating $P_1\times P_2 \times Q_1 \times Q_2$ parameters, which may cause overfitting or computational issues, especially when the number of parameters is larger than the sample size $N$. 

To reduce the number of free parameters while preserving the correlation structure in modeling tensor data, tensor decomposition techniques have been widely applied \citep{kolda2009tensor}. 
The two most commonly-used tensor decomposition methods are the   PARAFAC/CANDECOMP (CP) decomposition \citep{harshman1970foundations} and Tucker decomposition \citep{tucker1966some}. 
The CP decomposition reconstructs a tensor as a linear combination of rank-1 tensors, each one of which is represented as the outer product of a number of vectors. On the other hand, 
the Tucker decomposition 
factorizes a tensor into a small core tensor and a set of matrices along each dimension. 
Both decomposition methods are able to  reduce model dimensionality to a manageable size and make parameter estimation more efficient. 
Compared to CP decomposition, Tucker decomposition allows a more flexible correlation structure processed by the core tensor and the freedom  in choosing different orders, making it useful in estimating data with skewed dimensions \citep{li2013tucker}. In fact, CP decomposition is a special case of Tucker decomposition with the core tensor being superdiagonal. 

There is a rich literature on regression methods treating tensors as  either predictors or responses in both frequentist and Bayesian statistics. 
\cite{guo2012tensor} and \cite{zhou2013tensor} proposed tensor regression models to predict scalar outcomes from tensor predictors by assuming that the coefficient tensor has a low rank CP decomposition.  
\cite{li2013tucker} later extended the framework by employing Tucker decomposition for the coefficient tensor, and demonstrated that Tucker decomposition is more suitable to deal 
with tensor predictors of skewed dimensions and gains better  accuracy in neuroimaging data analysis. \cite{guhaniyogi2017bayesian} proposed a Bayesian approach to regression with a scalar response on tensor predictors by developing a multiway Dirichlet generalized double Pareto prior on tensor margins after applying CP decomposition to the coefficient tensor. \cite{miranda2018tprm} developed a Bayesian tensor partition regression model using  a generalized linear model with a sparse inducing normal mixture prior to learn the relationship between a matrix response (clinical outcomes) and a tensor predictor (imaging data). 
\cite{li2017parsimonious} proposed a parsimonious regression model with tensor response and vector predictors adopting a generalized sparsity principle based on Tucker decomposition. 
To detect neuronal activation in fMRI experiments,  \cite{guhaniyogi2021bayesian} developed a Bayesian regression approach with a tensor response on scalar predictors by introducing a novel multiway stick breaking shrinkage prior distribution on tensor structured regression coefficients.

There exist many scientific applications that require methods   for predicting a tensor response from another tensor predictor. 
One typical example in fMRI studies is to detect brain regions activated by an external stimulus or condition \citep{zhang2015bayesian}. \cite{hoff2015multilinear} proposed a tensor-on-tensor bilinear regression framework to handle a special case where the tensor predictor has the same dimension as the tensor response making use of Tucker decomposition.  \cite{billio2018bayesian} introduced a Bayesian tensor autoregressive model to tackle tensor-on-tensor regression, and used CP decomposition to provide parsimonious parametrization. 
\cite{lock2018tensor} proposed to predict a tensor response from another tensor predictor by assuming that the coefficient tensor has a low-rank CP factorization. 
\cite{gahrooei2020multiple} extended the work of \cite{lock2018tensor} to allow multiple tensor inputs under the Tucker decomposition framework. 

Despite advances in methods development for dealing with tensor data, there are some limitations in the aforementioned methods. First, tensor-on-tensor regression methods based on CP decomposition, e.g., \cite{lock2018tensor}, require both the response tensor and the predictor tensor to have the same rank in CP decomposition, making them restrictive when the response and predictor tensors have different ranks. Second, the rank in CP decomposition and the dimension of the core tensor in Tucker decomposition (i.e., model dimension) are essential for statistical inference in tensor-on-tensor regression models.  However, they are either assumed known or estimated via cross-validation \citep{gahrooei2020multiple} or some model selection criteria, such as Bayesian information criterion \citep{guhaniyogi2021bayesian}. To our best knowledge, there is no existing method that can simultaneously estimate the model dimension and parameters. 

In this paper, we develop a novel Bayesian approach for  tensor-on-tensor regression based on Tucker decomposition of the coefficient tensor. The main contributions of this work are threefold. First, our Bayesian framework is built upon the flexible Tucker decomposition so that the response tensor and the predictor tensor can have different dimensions in the core tensor. Second, we propose an efficient Markov chain Monte Carlo (MCMC) algorithm to simultaneously estimate the model dimension (the dimension of the core tensor) and parameters. The resulting posterior inference naturally offers us characterization of uncertainty in parameter estimation and prediction. Third, as an alternative to MCMC, we develop an ultra-fast computing algorithm, in which the maximum a posteriori (MAP) estimators for parameters are computed and meanwhile the  dimension of the core tensor is optimized via a simulated annealing (SA) algorithm \citep{kirkpatrick1983optimization}.  

The rest of the article is organized as follows. We start with introducing some preliminaries in section \ref{sec:pre}. section \ref{sec:BayTensor} describes the proposed Bayesian tensor-on-tensor regression model. We develop an efficient MCMC algorithm to  simultaneously estimate the model dimension and parameters in section \ref{sec:inference}. An optimization-based ultra-fast computational algorithm for inference is described in section \ref{sec:fast}.  section \ref{sec:simu} evaluates the proposed approach via simulation studies and comparisons to  alternative methods. 
section \ref{sec:real} provides real data analyses on facial imaging data and 3D motion data. section \ref{sec:dis} concludes with a discussion. 

\section{Preliminaries}	
\label{sec:pre}
\subsection{Notations}

We begin with introducing notations and operations that will be used throughout the paper. We use uppercase blackboard bold characters ($\mathbb{X}$) to denote tensors, bold uppercase characters ($\mathbf{X}$) to denote matrices, and bold lowercase characters ($\mathbf{a}$) to denote vectors. 
The\textit{ order} of a tensor is the number of dimensions. For example, $\mathbb{X}\in \Rbb^{I_1\times I_2 \times \cdots \times I_N}$ denotes an \textit{N}th order tensor, where $I_n$ denotes the dimension of the \textit{n}th mode, $n=1, \dots, N$. 
The \textit{i}th entry of a vector $\bm{a}$ is denoted as $a_i$; the element $(i,j)$ of a matrix $\mathbf{X}$ is denoted as $X_{ij}$; and the entries of a tensor are defined by indices enclosed in square brackets: $\mathbb{X}_{[i_1,\cdots,i_N]}$, where $i_n\in \{1,\cdots,I_n\}$ for $n\in \{1,\cdots N\}$. 
The $n$th element in a sequence of matrices or vectors is denoted by a subscript in parenthesis. For example, $\mathbf{X}_{(n)}$ denotes the $n$th matrix in a sequence of matrices, and $\mathbf{x}_{(n)}$ denotes the $n$th vector in a sequence of vectors.


The \textit{vectorization} of a tensor $\mathbb{X}\in \Rbb^{I_1\times I_2 \times \cdots \times I_N}$ transforms an \textit{N}th order tensor into a column vector \textit{vec}$ \mathbb{X}$ such that  the entry  $\mathbb{X}_{[i_1,\cdots,i_N]}$ maps to the \textit{j}th entry of \textit{vec}$ \mathbb{X}$, that is 
\begin{equation}
	\label{eq:vectorize}
	\mathbb{X}_{[i_1,\cdots,i_N]} = \textit{vec}\mathbb{X}_j\,,
\end{equation}where $j=1+\sum_{k=1}^{N}(i_k-1)\prod_{l=1}^{k-1}I_l$.  Similarly,  $\textit{vec}\mathbf{X}$ is used to denote the \textit{vectorization} of a matrix $\mathbf{X} \in \Rbb^{I_1\times I_2}$ when  $N = 2$ in \eqref{eq:vectorize}.
\textit{Matricization}, also known as unfolding, is the process of transforming a tensor into a matrix. 
The mode-$n$ matricization of a tensor  $\mathbb{X}\in \Rbb^{I_1\times I_2 \times \cdots \times I_N}$ is denoted by $\mathbb{X}_{(n)}\in \Rbb^{I_n\times J}$ where $J=\prod_{k\neq n}I_k$. The entry $\mathbb{X}_{[i_1,\cdots,i_N]}$ of $\mathbb{X}$ maps to the $(i_n,j)$ element of the resulting matrix $\mathbb{X}_{(n)}$,  where
$$j=1+\sum_{\substack{k=1 \\ k\neq n}}^{N}(i_k-1)J_k \quad \text{with } \quad J_k=\prod_{\substack{l=1 \\ l\neq n}}^{k-1}I_l.$$

A more general treatment of the tensor matricization is defined as follows. Let $\mathcal{R}=\{r_1,\cdots,r_L\}$ and $\mathcal{C}=\{c_1,\cdots,c_M\}$ be  two sets of indices such that $\mathcal{R}\cup \mathcal{C}=\{1,\cdots,N\}$ and $\mathcal{R} \cap \mathcal{C}=\emptyset$. Then the matricized tensor can be specified by $\mathbb{X}_{(\mathcal{R}\times \mathcal{C})}\in \Rbb^{J\times K}$, where $J=\prod_{n\in \mathcal{R}}I_n$ and  $K=\prod_{n\in \Ccal}I_n.$ And the entry $\mathbb{X}_{[i_1,\cdots,i_N]}$ maps to the $(j,k)$ element of the matrix $\mathbb{X}_{(\mathcal{R}\times \mathcal{C})}$, that is 
\begin{equation}
	\label{eq:matricize}
	\mathbb{X}_{[i_1,\cdots,i_N]} = \left(\mathbb{X}_{(\mathcal{R}\times \mathcal{C})}\right)_{jk} \,,
\end{equation}
where $$j=1+\sum_{l=1}^{L}\left[(i_{r_l}-1)\prod_{l'=1}^{l-1}I_{r_{l'}}\right] \,,$$
and 
$$
k=1+\sum_{m=1}^{M}\left[(i_{c_m}-1)\prod_{m'=1}^{m-1}I_{c_{m'}}\right]\,.
$$

The Kronecker product of matrices $\mathbf{U}\in \Rbb^{I\times J}$, and $\mathbf{V}\in \Rbb^{K\times L}$ is denoted by $\mathbf{U}\otimes \mathbf{V}$ with the detailed definition and properties shown in \ref{sec:kro}.
The product of a tensor and a matrix in mode $n$ is defined as the \textit{\textit{n}-mode product}. The \textit{n}-mode product of $\mathbb{X}\in \Rbb^{I_1\times I_2 \times \cdots \times I_N}$ with a matrix $\mathbf{U} \in \Rbb^{J\times I_n}$ is denoted by $\mathbb{X}\times_n \mathbf{U}$,  resulting in a new tensor 
$\mathbb{Y} \in \Rbb^{I_1\times \cdots \times I_{n-1} \times J \times I_{n+1} \times  \cdots \times I_N}$ where the $[i_1,\cdots i_{n-1},j,i_{n+1},\cdots i_N]$ entry 
is defined by $$\mathbb{Y}_{[i_1,\cdots, i_{n-1},j,i_{n+1},\cdots, i_N]}=\sum_{i_n=1}^{I_n}\mathbb{X}_{[i_1,\cdots,i_N]}U_{ji_n}.$$ An important fact regarding the \textit{n}-mode product is that 
given matrices $\mathbf{U}\in \Rbb^{J_1\times I_n}$, $\mathbf{V}\in \Rbb^{J_2\times I_m}$ with $m\neq n$, and tensor $\mathbb{X}\in \Rbb^{I_1\times I_2 \times \cdots \times I_N}$, then 
$$\mathbb{X}\times_n \mathbf{U} \times_m \mathbf{V}=(\mathbb{X}\times_n \mathbf{U})\times_m \mathbf{V}=(\mathbb{X}\times_m \mathbf{V})\times_n \mathbf{U}.$$

For two tensors $\mathbb{X}\in \Rbb^{I_1\times \cdots \times I_N \times P_1 \times \cdots \times P_L}$, and $\mathbb{Y}\in \Rbb^{P_1 \times \cdots \times P_L \times J_1 \times \cdots \times J_M}$, the \textit{contracted tensor product} $\langle\mathbb{X},\mathbb{Y} \rangle_L$ is defined as 
$$\mathbb{Z}=\langle\mathbb{X},\mathbb{Y} \rangle_L \in \Rbb^{I_1\times \cdots \times I_N\times J_1\times \cdots \times J_M}$$
with \begin{align*}
	& \mathbb{Z}_{[i_1,\cdots,i_N,j_1,\cdots,j_M]}= \\
	& \sum_{p_1=1}^{P_1}\cdots \sum_{p_L=1}^{P_L}\mathbb{X}_{[i_1,\cdots,i_N,p_1,\cdots,p_L]}\mathbb{Y}_{[p_1,\cdots,p_L,j_1,\cdots,j_M]}.
\end{align*}
It can be shown that for two matrices $\mathbf{U}\in \Rbb^{I\times P}$ and $\mathbf{V}\in \Rbb^{P\times J}$, the contracted product $\langle\mathbf{U},\mathbf{V} \rangle_1$ is equivalent to the standard matrix product $\mathbf{UV}$. Therefore, the contracted product of two tensors can be regarded as  an extension of the usual matrix product to higher-order operands. 

\subsection{Tucker decomposition}

The proposed Bayesian tensor-on-tensor model is built upon Tucker decomposition \citep{tucker1966some}, which decomposes a tensor $\mathbb{B}\in \Rbb^{I_1\times I_2 \times \cdots \times I_N}$ into a core tensor $\mathbb{G}$ and a set of factor matrices $\mathbf{A}_{(n)}$, $n=1, \dots, N$, denoted by 
$$\mathbb{B}=[\![\mathbb{G};\mathbf{A}_{(1)},\mathbf{A}_{(2)},\cdots,\mathbf{A}_{(N)}]\!].$$
Or equivalently, 
$$\mathbb{B}=\mathbb{G}\times_1 \mathbf{A}_{(1)}\times_2 \mathbf{A}_{(2)}\cdots\times_N\mathbf{A}_{(N)},$$
with $\mathbb{G}\in \Rbb^{J_1\times \cdots \times J_N}$ being the core tensor and $\mathbf{A}_{(n)}\in \Rbb^{I_n\times J_n}$ being the factor matrix in mode $n$,  for $n=1,\cdots, N$, forming a sequence of matrices. The order  of $\mathbb{G}$ can be the same as the order of $\mathbb{B}$, but more often, we are interested in compressing the information of $\mathbb{B}$  to a smaller size of $\mathbb{G}$ than $\mathbb{B}$.   CP decomposition \citep{harshman1970foundations} is a special case of Tucker decomposition wherein the core tensor is superdiagonal. 

\begin{figure}[H]
	\centering
	\includegraphics[width=0.9\textwidth]{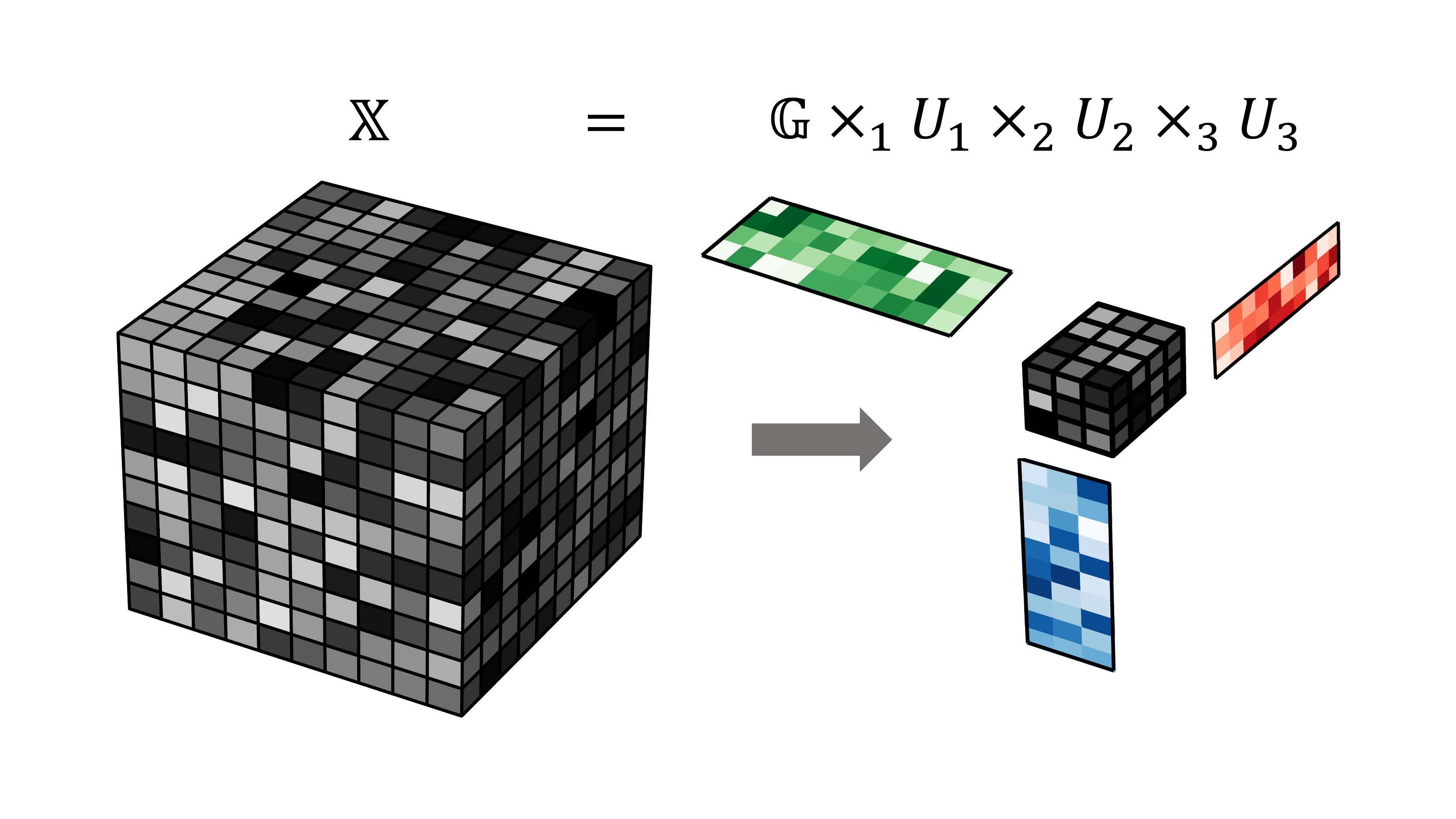}
	\caption{Illustration of Tucker decomposition. Here, the  core tensor is of dimension $(4,3,3)$.  }
	\label{tucker}
\end{figure}

\section{A Bayesian Tensor-on-Tensor Regression Model}
\label{sec:BayTensor}
Our task is to predict a tensor response $\mathbb{Y}\in \Rbb^{N\times Q_1\times \cdots \times Q_M}$ from a tensor predictor $\mathbb{X}\in \Rbb^{N\times P_1\times \cdots \times P_L}$.  We propose a Bayesian tensor-on-tensor regression framework by extending the standard multivariate linear regression model from matrices to tensors: 
\begin{equation}
	\label{eq:tensor_reg1}
	\mathbb{Y}=\langle\mathbb{X},\mathbb{B} \rangle_L+\mathbb{E},
\end{equation}
where $\mathbb{B}\in \Rbb^{P_1\times \cdots \times P_L \times Q_1\times \cdots \times Q_M}$ denotes the coefficient tensor, and each element of $\mathbb{E}\in \Rbb^{N\times Q_1\times \cdots \times Q_M}$ is assumed to follow $N(0, \sigma^2)$ independently. 
The first $L$ modes of $\mathbb{B}$ contract the dimensions of $\mathbb{X}$ and the last $M$ modes of $\mathbb{B}$ match the modes of $\mathbb{Y}$. For each of the $N$ observations, there are a total of  $(\prod_{m=1}^{M}Q_m)$ responses and $(\prod_{l=1}^{L}P_l)$ predictors. 
The model can be reformulated into a matrix form as follows,
\begin{equation}
	\label{eq:tensor_reg2}
	\mathbb{Y}_{(1)}=\mathbb{X}_{(1)}\mathbb{B}_{(\mathcal{P}\times \mathcal{Q})}+\mathbb{E}_{(1)},
\end{equation}
where $\mathcal{P}=\{1,\cdots,L\}$, $\mathcal{Q}=\{L+1,\cdots,L+M \}$,
and each row of $\mathbb{E}_{(1)}$ independently follows $N(0, \sigma^2 \mathbf{I}_d)$ with  $\mathbf{I}_d$ being an identity matrix of dimension $(\prod_{m=1}^{M}Q_m)$. 
From the equivalence of  \eqref{eq:tensor_reg1} and \eqref{eq:tensor_reg2}, it is clear that our model \eqref{eq:tensor_reg1} supports linear relations between responses and predictors for each observation. 

If $\mathbb{B}$ is unconstrained, the estimation of $\mathbb{B}$ can be obtained by conducting separate ordinary least squares (OLS) regressions  for each of the $\prod_{m=1}^{M}Q_m$ responses in $\mathbb{Y}_{(1)}$ over $\mathbb{X}_{(1)}$ by equation \eqref{eq:tensor_reg2}  in the frequentist framework. 
However, the solution is not well-defined if  the number of observations $N$ is less than the number of responses $\prod_{m=1}^{M}Q_m$. Even if the solution is well-defined, separate OLS of equation \eqref{eq:tensor_reg2} does not consider the correlation structure 
within the response $\mathbb{Y}$ and predictor $\mathbb{X}$ and between them.  Moreover, the total number of parameters in this case is  $\prod_{l=1}^{L}P_l\prod_{m=1}^{M}Q_m$, which  can be computationally challenging due to its gigantic size. 

In our model, we assume that $\mathbb{B}$ follows the Tucker decomposition defined by
\begin{equation}
	\label{eq:tucker_1}
	\mathbb{B}=[\![\mathbb{G};\mathbf{U}_{(1)},\cdots,\mathbf{U}_{(L)},\mathbf{V}_{(1)},\cdots,\mathbf{V}_{(M)}]\!],
\end{equation}
where $\mathbb{G}$ denotes the core tensor of dimensions $R_1\times \cdots \times R_L\times S_1\times \cdots\times S_M$, $\mathbf{U}_{(l)} \in \Rbb^{P_l\times R_l}$ denotes the factor matrix corresponding to the mode $l$ in $\mathbb{B}$ for $l = 1,\cdots,L$, and $\mathbf{V}_{(m)} \in \Rbb^{Q_m\times S_m}$ denotes the factor matrix corresponding to the mode $L+m$ in $\mathbb{B}$ for $m = 1,\cdots, M$. 

We complete the proposed Bayesian tensor-on-tensor regression model by assigning priors to the core tensor $\mathbb{G}$, $\{\mathbf{U}_{(l)}\}_{l=1}^L$, $\{\mathbf{V}_{(m)}\}_{m=1}^M$, and $\sigma^2$.  
For the core tensor $\mathbb{G}$, we consider a normal prior, that is \textit{vec}$ \mathbb{G} \ \sim \ N(\bm{\mu}_G,\mathbf{\Sigma}_G)$ where $\mathbf{\Sigma}_G$ is diagonal. 
For factor matrices $\{\mathbf{U}_{(l)}\}_{l=1}^{L}$ and $\{\mathbf{V}_{(m)}\}_{m=1}^{M}$, taking $\mathbf{U}_{(l)}$ and $\mathbf{V}_{(m)}$ as examples, we assign normal priors for $vec\mathbf{U}^{(l)}$ and $vec\mathbf{V}^{(m)}$. That is, $vec\mathbf{U}_{(l)} \ \sim \ N(\bm{\mu}_{U_l},\mathbf{\Sigma}_{U_l})$, and
$vec\mathbf{V}_{(m)} \ \sim \ N(\bm{\mu}_{V_m}, \mathbf{\Sigma}_{V_m})$, where $\mathbf{\Sigma}_{U_l}$ and $\mathbf{\Sigma}_{V_m}$ are diagonal matrices.
Usually we choose $\bm{\mu}_{U_l}=\bm{\mu}_U$, $\bm{\mu}_{V_m}=\bm{\mu}_V$,  $\mathbf{\Sigma}_{U_l}=\mathbf{\Sigma}_{U}$, and $\mathbf{\Sigma}_{V_m}=\mathbf{\Sigma}_{V}$  for all $l=1,\cdots,L$ and $m = 1,\cdots,M$. 
Lastly we assign an inverse gamma prior distribution for 
$\sigma^2$:  $\sigma^2 \ \sim IG(\alpha,\beta)$.


\section{Posterior Inference}
\label{sec:inference}
We conduct posterior inference using Markov chain Monte Carlo (MCMC) simulations. Given the dimension $\bm{\theta}=(R_1,\dots,R_L,S_1,\cdots,S_M)$ of the core tensor, we update $\mathbf{U}_{(l)}$, $\mathbf{V}_{(m)}$, $\mathbb{G}$, and $\sigma^2$ using Gibbs sampling transition probabilities for posterior updates, the details of which will be given in section \ref{sec:mcmc1}. The posterior update of the core tensor dimension $\bm{\theta}$ is challenging since the dimensions of  $\mathbf{U}_{(l)}$, $\mathbf{V}_{(m)}$, and $\mathbb{G}$ change when $\bm{\theta}$ varies. A reversible jump (RJ) MCMC \citep{green1995reversible} algorithm is a natural choice for such a trans-dimensional update, however, it is difficult to construct a practicable RJ scheme due to the high-dimensionality of the problem. To address this challenge, we will develop an efficient Metropolis-Hastings (MH) algorithm to update $\bm{\theta}$ building upon the idea of fractional Bayes factor \citep{lee2016bayesian,o1995fractional} in section \ref{sec:mcmc2}.


\subsection{Posterior inference given the dimension of the core tensor}
\label{sec:mcmc1}
Given the dimension $\bm{\theta}= (R_1, \cdots, R_L,S_1, \cdots, S_M)$ of the core tensor $\mathbb{G}$,  we derive 
the full conditional posterior distributions  of $\{\mathbf{U}_{(l)}\}_{l=1}^{L}$, $\{\mathbf{V}_{(m)}\}_{m = 1}^{M}$, $\mathbb{G}$, and $\sigma^2$  in closed forms.  Without loss of generality, we first derive the full conditional posterior distribution of $\mathbf{U}_{(1)}$.  The full conditional posterior distributions of $\{\mathbf{U}_{(2)},\cdots,\mathbf{U}_{(L)}\}$ can be derived in the same manner. 

By properties of \textit{n}-mode product of tensor and Tucker decomposition, we have
$$\mathbb{B}=\mathbb{G}\times_2 \mathbf{U}_{(2)} \cdots \times_L \mathbf{U}_{(L)} \times_{L+1} \mathbf{V}_{(1)} \cdots \times_{L+M} \mathbf{V}_{(M)} \times_1 \mathbf{U}_{(1)}.$$ 
Let $\mathbb{B}_{(-)}$ denote $\mathbb{G}\times_2 \mathbf{U}_{(2)} \cdots \times_L \mathbf{U}_{(L)} \times_{L+1} \mathbf{V}_{(1)} \cdots \times_{L+M} \mathbf{V}_{(M)}$, then $\mathbb{B}_{(-)}\in \Rbb^{R_1\times P_2\times\cdots \times P_L \times Q_1 \times \cdots \times Q_M}$, and
$$\mathbb{B}=\mathbb{B}_{(-)}\times_1 \mathbf{U}_{(1)}.$$
We denote the contracted product of $\langle \mathbb{B}_{(-)},\mathbb{X}\rangle_{P_2,\cdots,P_N}$ by a new tensor called $\mathbb{C}$, then tensor $\mathbb{C}\in \Rbb^{R_1\times N \times P_1 \times Q_1\times \cdots Q_M}$. By tensor matricization of  $\mathbb{C}$ into $\mathbb{C}_{(\mathcal{R}\times \Ccal)} \in \Rbb^{N\prod_{m=1}^{M}Q_m \times R_1P_1}$, where $\mathcal{R}=\{N,Q_1,\cdots,Q_M\}$, and $\mathcal{C}=\{R_1,P_1\}$, we can rewrite our model \eqref{eq:tensor_reg1} as follows: 
\begin{equation}
	\label{eq:post_u1}
	vec\mathbb{Y}= \mathbb{C}_{(\mathcal{R}\times \Ccal)} \times \ vec \mathbf{U}_{(1)}+vec\mathbb{E}.
\end{equation}
The proof for equation \eqref{eq:post_u1} is given in Appendix \ref{sec:post_u1}. 


\begin{figure}[H]
	\centering
	\begin{subfigure}{0.9\textwidth}
		\includegraphics[width=\textwidth]{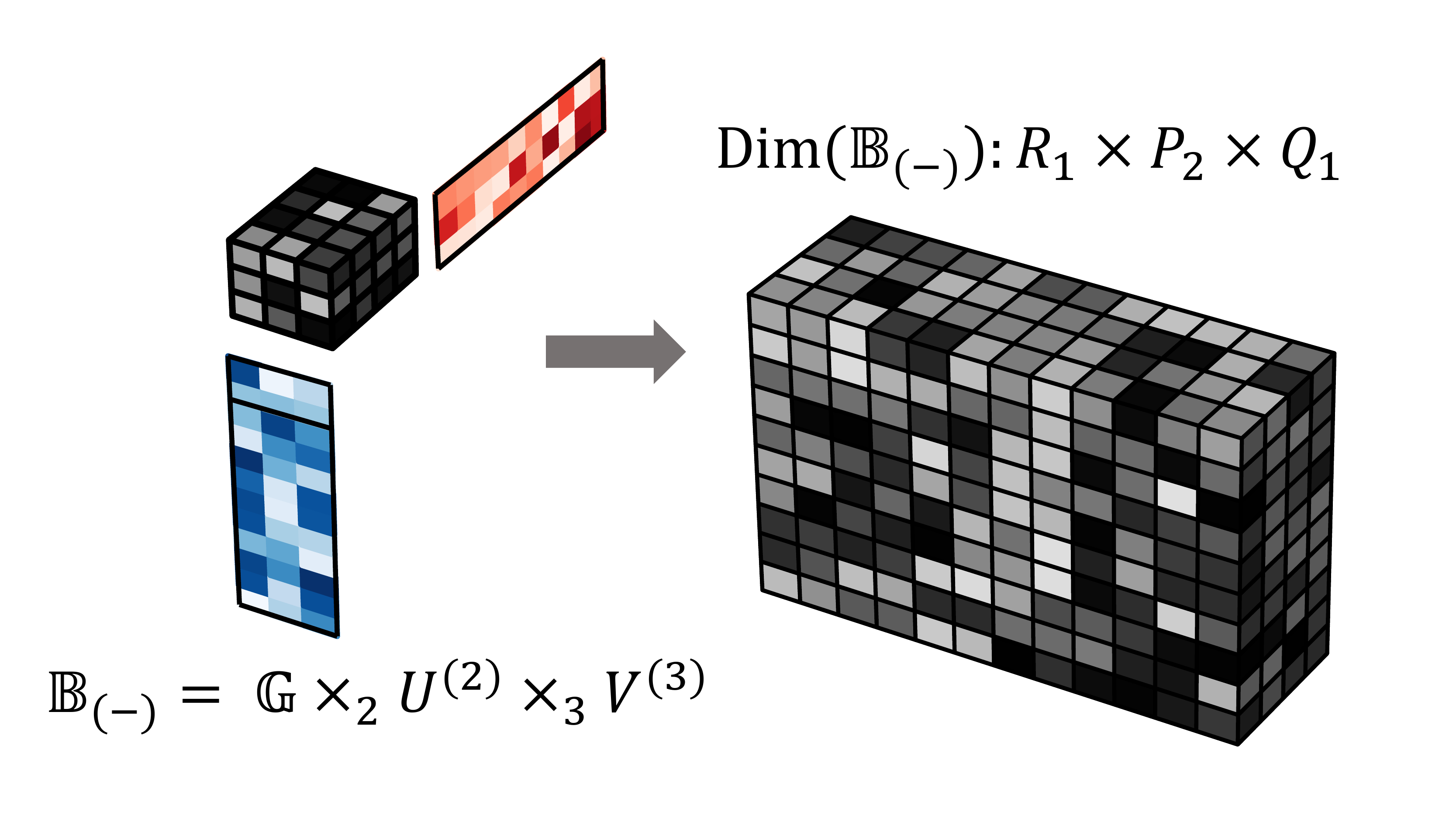}
		\caption{Calculate $\mathbb{B}_{(-)}$.}
	\end{subfigure}
	\begin{subfigure}{0.9\textwidth}
		\includegraphics[width=\textwidth]{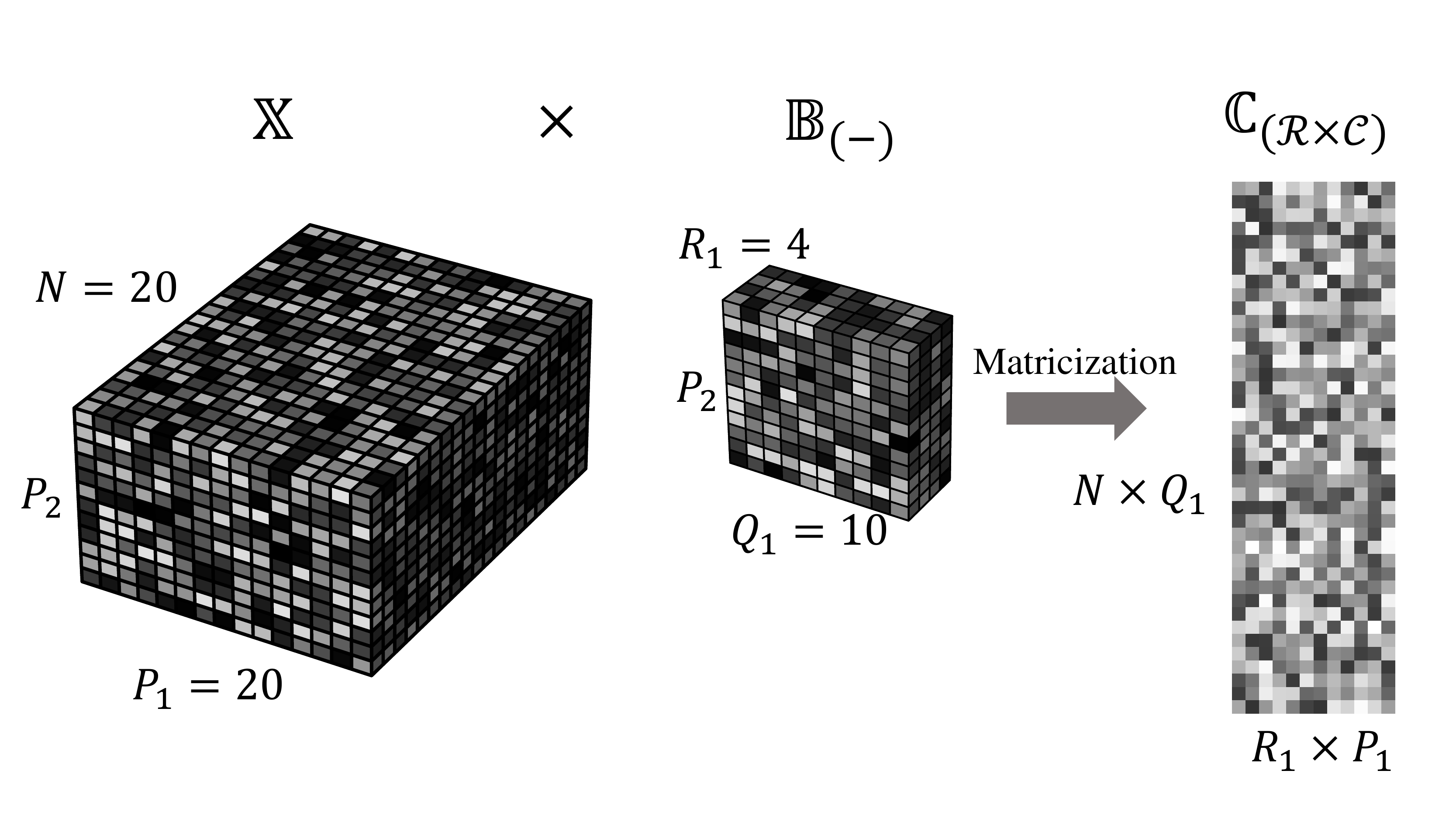}
		\caption{Calculate $\mathbb{C}_{(\mathcal{R} \times \Ccal)}$.}
	\end{subfigure}
	\caption{Illustration of updating $\mathbf{U}_{(1)}$.}
	\label{fig:fig1}
\end{figure}

Following \eqref{eq:post_u1}, we can easily derive that the  full conditional posterior distribution of $vec\mathbf{U}_{(1)}$ is normally distributed:  
\begin{equation}
	\label{eq:U_posterior}
	p(vec\mathbf{U}_{(1)} \mid vec\mathbb{Y},\mathbb{X},\sigma^2, \mathbf{V}_{(m)}, \mathbf{U}_{(l)} \ l\neq 1)\sim \ N(\bm{\mu}'_U, \mathbf{\Sigma}'_U),
\end{equation} where
\begin{align*}
	\mathbf{\Sigma}'_U & = \left(\frac{\mathbb{C}_{(\mathcal{R}\times \Ccal)}^T\mathbb{C}_{(\mathcal{R}\times \Ccal)}}{\sigma^2}+\mathbf{\Sigma}_U^{-1} \right)^{-1} \,,\\
	\bm{\mu}_U^{'} &= \mathbf{\Sigma}'_U \left(\frac{\mathbb{C}_{(\mathcal{R}\times \Ccal)}^Tvec\mathbb{Y}}{\sigma^2}+\mathbf{\Sigma}_U^{-1}\bm{\mu}_U\right) \,.
\end{align*}
Figure \ref{fig:fig1} presents an illustration of updating $\mathbf{U}_{(1)}$.

We then derive the conditional distributions of 
$\mathbf{V}_{(m)}$ given $\sigma^2$, $\{\mathbf{U}_{(l)}\}_{l=1}^{L}$, $\mathbf{V}_{(k)}$ for $k\neq m$, and $\mathbb{G}$. Without loss of generality, we derive the full conditional posterior distribution of $\mathbf{V}_{(1)}$ below. 

Denote the contracted product of the tensor  $\mathbb{G}\times_1 \mathbf{U}_{(1)} \cdots \times_L \mathbf{U}_{(L)} \times_{L+2} \mathbf{V}_{(2)} \cdots \times_{L+M} \mathbf{V}_{(M)}$ and the tensor $\mathbb{X}$ by a new tensor $\mathbb{D}$, where $\mathbb{D}\in \Rbb^{N\times S_1\times Q_2\times \cdots \times Q_M}$. We then matricize $\mathbb{D}$ into a matrix $\mathbb{D}_{(\mathcal{R}\times \Ccal)} \in \Rbb^{N\prod_{m=2}^{M}Q_m \times S_1}$ and write 
\begin{equation}
	\label{eq:post_v1}
	\mathbb{{Y}}_{(2)}=\mathbf{V}_{(1)}\times (\mathbb{D}_{(\mathcal{R}\times \Ccal)})^T+\mathbb{E}_{(2)},
\end{equation}
where $\mathbb{{Y}}_{(2)}\in \Rbb^{Q_1\times N\prod_{m=2}^{M}Q_m}$ is the mode-$2$ matricization of tensor $\mathbb{Y}$. The proof of equation \eqref{eq:post_v1} can be found in Appendix \ref{sec:post_v1}.
Let $\mathbf{\tilde{Y}}=\mathbb{{Y}}^T_{(2)}$. Given that $\mathbf{V}_{(1)}$ follows a normal distribution with a diagonal covariance matrix, 
we can rewrite \eqref{eq:post_v1} as 
\begin{equation*}
	vec\tilde{\mathbf{Y}} = \left(\mathbf{I}_{Q_1}\otimes\mathbb{D}_{(\mathcal{R}\times \Ccal)}\right) \times vec\mathbf{V}_{(1)}^T + vec\left(\mathbb{E}_{(2)}\right)^T,
\end{equation*}
where $\mathbf{I}_{Q_1}$ denotes an identity matrix of size $Q_1$. Given that the prior distribution of $vec\mathbf{V}_{(1)}$ is a normal $N(\bm{\mu}_{V},\bm{\Sigma}_V)$ with diagonal $\bm{\Sigma}_V$, the prior distribution of $vec\mathbf{V}_{(1)}^{T}$ is also a normal distribution $N(\tilde{\bm{\mu}}_V, \tilde{\bm{\Sigma}}_V)$ with a diagonal covariance matrix.
Then the full conditional posterior distribution of $vec{\mathbf{V}_{(1)}}^T$ is normally distributed:
\begin{equation}
	\label{eq:V_posterior}
	p(vec\mathbf{V}_{(1)}^T \mid vec\tilde{\mathbf{Y}},\mathbb{X},\sigma^2, \mathbf{U}_{(l)}, \mathbf{V}_{(m)} \ m\neq 1)\sim \ N(\tilde{\bm{\mu}}_V^{'}, \tilde{\mathbf{\Sigma}}_V^{'}),
\end{equation}
where
\begin{equation}
	\begin{split}
		\tilde{\mathbf{\Sigma}}_{V}^{'}&=\left(\frac{\left(\mathbf{I}_{Q_1}\otimes\mathbb{D}_{(\mathcal{R}\times \Ccal)}\right)^T\left(\mathbf{I}_{Q_1}\otimes\mathbb{D}_{(\mathcal{R}\times \Ccal)}\right)}{\sigma^2}+\tilde{\mathbf{\Sigma}}_V^{-1}\right)^{-1}, \\
		\tilde{\bm{\mu}}_{V}^{'}&=\tilde{\mathbf{\Sigma}}_{V}^{'} \left(\frac{\left(\mathbf{I}_{Q_1}\otimes\mathbb{D}_{(\mathcal{R}\times \Ccal)}\right)^T vec\mathbf{\tilde{Y}}}{\sigma^2}+\tilde{\mathbf{\Sigma}}_V^{-1}\tilde{\bm{\mu}}_V\right).
	\end{split}
\end{equation}

The posterior distribution of the core tensor $\mathbb{G}$ is more complex than the posterior of $\mathbf{U}_l's$ and $\mathbf{V}_m's$. First, we have 
\begin{equation}
	\label{eq:post_g1}
	\begin{split}
		\mathbb{Y}_{(1)} & =\mathbb{X}_{(1)}\mathbb{B}_{(\mathcal{P}\times \mathcal{Q})}+\mathbb{E}_{(1)}  \\
		&=\mathbb{X}_{(1)}\left(\mathbf{U}_{(L)}\otimes \cdots \otimes \mathbf{U}_{(1)}\right)\mathbb{G}_{(\mathcal{R}\times \mathcal{C})}
		\\
		&\times \left(\mathbf{V}_{(M)}\otimes \cdots \otimes \mathbf{V}_{(1)}\right)^T +\mathbb{E}_{(1)}. 
	\end{split}
\end{equation}
Let $\mathbf{U}$ denote $\left(\mathbf{U}_{(L)}\otimes \cdots \otimes \mathbf{U}_{(1)}\right)$, and $\mathbf{V}$ denote $\left(\mathbf{V}_{(M)}\otimes \cdots \otimes \mathbf{V}_{(1)}\right)$, we have 
\begin{equation}
	\label{eq:post_g2}
	\mathbb{Y}_{(1)}=(\mathbb{X}_{(1)}\mathbf{U})\mathbb{G}_{(\mathcal{R}\times \mathcal{C})}\mathbf{V}^T+\mathbb{E}_{(1)}.
\end{equation}
If we further let $\dbtilde{\mathbf{Y}}$ denote $\mathbb{Y}_{(1)}\mathbf{V}(\mathbf{V}^T\mathbf{V})^{-1}$, and $\dbtilde{\mathbf{E}}$ denote $\mathbb{E}_{(1)}\mathbf{V}(\mathbf{V}^T\mathbf{V})^{-1}$, we can rewrite \eqref{eq:post_g2} as
\begin{equation*}
	\dbtilde{\mathbf{Y}} = (\mathbb{X}_{(1)}\mathbf{U})\mathbb{G}_{(\mathcal{R}\times \mathcal{C})} + \dbtilde{\mathbf{E}} \,,
\end{equation*}
or 
\begin{equation*}
	vec\dbtilde{\mathbf{Y}} = (\mathbf{I}_S\otimes (\mathbb{X}_{(1)}\mathbf{U})) vec\mathbb{G}_{(\mathcal{R}\times \mathcal{C})} + vec\dbtilde{\mathbf{E}} \,,
\end{equation*}
where $\mathbf{I}_S$ denotes an $S\times S$ identity matrix with $S=\prod_{m=1}^{M}S_m$, and $vec\dbtilde{\mathbf{E}}$ is normally distributed with mean $\mathbf{0}$ and block diagonal covariance matrix $\sigma^2(\mathbf{V}^T\mathbf{V})^{-1}\otimes \mathbf{I}_{N}$.
Then $vec \dbtilde{\mathbf{Y}} \ \sim \ N(\bm{\mu}_{\dbtilde{Y}}, \mathbf{\Sigma}_{\dbtilde{Y}})$ with
\begin{equation*}
	\begin{split}
		\bm{\mu}_{\dbtilde{Y}}&=(\mathbf{I}_S\otimes (\mathbb{X}_{(1)}\mathbf{U}))\times vec \mathbb{G}_{(\mathcal{R}\times \mathcal{C})} \,, \\
		\mathbf{\Sigma}_{\dbtilde{Y}}&=\sigma^2(\mathbf{V}^T\mathbf{V})^{-1}\otimes \mathbf{I}_{N} \,.
	\end{split}
\end{equation*}
Given that the prior distribution of \textit{vec}$ \mathbb{G}$ is $N(\bm{\mu}_G,\mathbf{\Sigma}_G)$ with a diagonal covariance $\mathbf{\Sigma}_G$, $vec \mathbb{G}_{(\mathcal{R}\times \mathcal{C})}$ is also normally distributed with $\tilde{\bm{\mu}}_G$ and diagonal covariance $\tilde{\bm{\Sigma}}_G$ by rearranging elements of $\bm{\mu}_G$ and $\mathbf{\Sigma}_G$.
Then the full conditional posterior distribution of $vec \mathbb{G}_{(\mathcal{R}\times \mathcal{C})}$ is a normal distribution with 
\begin{equation}
	\label{eq:post_g3}
	\begin{split}
		&\tilde{\bm{\mu}}_{G}^{'} =\tilde{\mathbf{\Sigma}}_G^{'} \left((\mathbf{I}_S\otimes (\mathbb{X}_{(1)}\mathbf{U}))^T(\mathbf{\Sigma_{\dbtilde{Y}}})^{-1} vec\mathbf{\dbtilde{Y}}+\tilde{\mathbf{\Sigma}}_G^{-1}\tilde{\bm{\mu}}_G\right) \,, \\
		& \tilde{\mathbf{\Sigma}}_G^{'} =\\ 
		& \left((\mathbf{I}_S\otimes (\mathbb{X}_{(1)}\mathbf{U}))^T  (\mathbf{\Sigma_{\dbtilde{Y}}})^{-1}(\mathbf{I}_S\otimes (\mathbb{X}_{(1)}\mathbf{U}))+(\tilde{\mathbf{\Sigma}}_G)^{-1}\right)^{-1} \,.
	\end{split}
\end{equation}

Lastly, deriving the full conditional posterior distribution of $\sigma^2$ is straightforward:
\begin{equation}
	\label{eq:sigma2_Posterior}
	p(\sigma^2 \mid \mathbb{Y},\mathbb{X},  \{\mathbf{U}_{(l)}\}_{l=1}^{L}, \{\mathbf{V}_{(m)}\}_{m = 1}^{M}, \mathbb{G}) \ \sim \ IG(\alpha', \beta'),
\end{equation}
where $\alpha'=\alpha+\frac{NQ}{2}$,  $\beta'=\beta+\frac{\norm{\mathbb{Y}-\langle \mathbb{X,B}\rangle_L}^2_F}{2}$ with $\mathbb{B}$ defined in \eqref{eq:tucker_1}, 
and $Q=\prod_{m=1}^{M}Q_m$.

\subsection{Updating the model dimension}
\label{sec:mcmc2}

In this subsection, we show how to simultaneously update the dimension of the core tensor and estimate model parameters. Denote $\bm{\theta}=(R_1,\dots,R_L,S_1,\cdots,S_M)$, and we assign a prior distribution $\pi(\bm{\theta})$ to $\bm{\theta}$. Since conditional posterior distribution of ${\bm \theta}$ is not in closed form,  we employ a trans-dimensional Metropolis-Hastings (MH) sampler to update  ${\bm \theta}$. The most challenging task is to design a good proposal distribution that can result in a reasonable acceptance rate given the fact that the dimensions of $\{\mathbf{U}_{(l)}\}_{l=1}^{L}$, $\{\mathbf{V}_{(m)}\}_{m=1}^{M}$, and $\mathbb{G}$ change when $\bm{\theta}$ varies. 

To address the challenge, we construct our proposal distribution building upon the idea of fractional Bayes factor  \citep{o1995fractional, lee2016bayesian}. Assuming that at iteration $t-1$ of the MCMC sampler  $\bm{\theta}^{(t-1)}=(R^{(t-1)}_{1},\cdots,R^{(t-1)}_{L},S^{(t-1)}_{1}\cdots,S^{(t-1)}_{M})$, 
at iteration $t$ we generate a candidate $\tilde{\bm{\theta}}$ from the ``neighbor" of $\bm{\theta}^{(t-1)}$  defined as $O(\bm{\theta}^{(t-1)}):=\{\tilde{\bm{\theta}} \in \bm{\Theta}: \lVert\bm{\theta}^{(t-1)}-\tilde{\bm{\theta}}\rVert_{L_1}=1\}$, where $\bm{\Theta}$ is the parameter space for ${\bm \theta}$. In this work, we propose to generate $\tilde{\bm{\theta}}$ uniformly over all candidates  in $O(\bm{\theta}^{(t-1)})$, denoted by  $q(\tilde{\bm{\theta}}\mid \bm{\theta}^{(t-1)})$.
To calculate the acceptance rate of the proposed $\tilde{\bm{\theta}}$ in the MH step, we denote $\bm{\xi} = \left(\{\mathbf{U}_{(l)}\}_{l=1}^{L}, \{\mathbf{V}_{(m)}\}_{m=1}^{M}, \mathbb{G}, \sigma^2 \right)$, and 
write the likelihood function as the multiplication of two parts:
\begin{equation*}
p(\mathbb{Y}\mid \bm{\xi},\bm{\theta})
=\underbrace{p(\mathbb{Y} \mid  \bm{\xi},\bm{\theta} )^{b}}_{\mathrm{training}}
\times \underbrace{p(\mathbb{Y} \mid  \bm{\xi},\bm{\theta} )^{(1-b)}}_{\mathrm{testing}},
\end{equation*}
where $b$ is small and $0<b<1$. The key idea here is to utilize a fraction $b$ of the data as the training data to propose new parameters $\bm{\xi}$ associated with $\tilde{\bm{\theta}}$ so that the new values can be accepted with a reasonable acceptance probability.  In particular, 
we update $\bm{\theta}^{(t)}$ as follows:
\begin{itemize}
\item Generate $\tilde{\bm{\theta}}$ from $q(\cdot \mid \bm{\theta}^{(t-1)})$.
\item Generate $\tilde{\bm{\xi}}$ from a distribution proportional to 
\[
p(\mathbb{Y}\mid \tilde{\bm{\xi}},\tilde{\bm{\theta}}) ^b \times p(\tilde{\bm{\xi}}\mid \tilde{\bm{\theta}}) \,,
\]	
which is  the posterior of $\bm{\xi}$ based on the training portion conditional on $\tilde{\bm{\theta}}$. 
The detailed conditional posterior distributions are shown in Appendix \ref{sec:post_givenb}.
\item Generate $\dbtilde{\bm{\xi}}$ from a distribution proportional to 
\[
p(\mathbb{Y}\mid \dbtilde{\bm{\xi}},\bm{\theta}^{(t-1)}) ^b \times p(\dbtilde{\bm{\xi}}\mid \bm{\theta}^{(t-1)}) \,,
\]	which is  the posterior of $\bm{\xi}$ based on the training portion conditional on $\bm{\theta}^{(t-1)}$.
\item Accept $\tilde{\bm{\theta}}$ with probability  $\min\left(1, A(\tilde{\bm{\theta}}, \bm{\theta}^{(t-1)} \right)$ where
\end{itemize}
\begin{equation}
\label{eq:MH_1}
A(\tilde{\bm{\theta}}, \bm{\theta}^{(t-1)})  
= \frac{\pi(\tilde{\bm{\theta}})q(\bm{\theta}^{(t-1)}\mid \tilde{\bm{\theta}})}{\pi(\bm{\theta}^{(t-1)})q(\tilde{\bm{\theta}} \mid \bm{\theta}^{(t-1)})}  \times \underbrace{\frac{p(\mathbb{Y}\mid \tilde{\bm{\xi}}, \tilde{\bm{\theta}})^{(1-b)}}{p(\mathbb{Y}\mid \dbtilde{\bm{\xi}}, \bm{\theta}^{(t-1)})^{(1-b)}}}_{(*)} \,.
\end{equation}
The (*) part in equation \eqref{eq:MH_1} coincides with  the fractional Bayes factor given in \cite{o1995fractional}. The detailed proof is given in Appendix \ref{sec:Frac_Bayes}.
\begin{algorithm}[h]
\caption{MCMC Sampler}
\label{algo:one}
\begin{algorithmic}[1]
	\renewcommand{\algorithmicrequire}{\textbf{Input:}}
	\renewcommand{\algorithmicensure}{\textbf{Output:}}
	\STATE Input data $\mathbb{X},\mathbb{Y}$.
	\STATE Initialize the core tensor dimension $\bm{\theta}^{(0)}$.
	\FOR {$t=1,\cdots,T$}
	\STATE Propose $\tilde{\bm{\theta}}\in O(\bm{\theta}^{(t-1)})$ from $q(\cdot \mid \bm{\theta}^{(t-1)})$.
	\STATE Sample $\left(\{\tilde{\mathbf{U}}_{(l)}\}_{l=1}^{L}, \{\tilde{\mathbf{V}}_{(m)}\}_{m=1}^{M},\tilde{\mathbb{G}}, \tilde{\sigma}^{2}\right)$ from full posterior based on the training portion $b$ conditional on $\tilde{\bm{\theta}}$ according to Appendix \ref{sec:post_givenb}.
	\STATE Sample $\left(\{\dbtilde{\mathbf{U}_{(l)}}\}_{l=1}^{L}, \{\dbtilde{\mathbf{V}}_{(m)}\}_{m=1}^{M}, \dbtilde{\mathbb{G}}, \dbtilde{\sigma}^2\right)$ from full posterior based on the training portion $b$ conditional on $\bm{\theta}^{(t-1)}$ according to Appendix \ref{sec:post_givenb}.
	\STATE Given $\left(\{\tilde{\mathbf{U}}_{(l)}\}_{l=1}^{L}, \{\tilde{\mathbf{V}}_{(m)}\}_{m=1}^{M},\tilde{\mathbb{G}}, \tilde{\sigma}^{2}, \tilde{\bm{\theta}} \right)$ and $\left(\{\dbtilde{\mathbf{U}_{(l)}}\}_{l=1}^{L}, \{\dbtilde{\mathbf{V}}_{(m)}\}_{m=1}^{M}, \dbtilde{\mathbb{G}}, \dbtilde{\sigma}^2, \bm{\theta}^{(t-1)}\right)$, calculate the acceptance probability according to \eqref{eq:MH_1}.
	\STATE Update and save $\bm{\theta}^{(t)}$ given the acceptance probability.
	\STATE Sample from full conditional posterior distributions:
	\INDSTATE Sample $\{vec\mathbf{U}^{(t)}_{(l)}\}_{l=1}^{L}$ according to \eqref{eq:U_posterior}.
	\INDSTATE Sample $\{vec\left(\mathbf{V}^{T}_{(m)}\right)^{(t)}\}_{m=1}^{M}$ according to \eqref{eq:V_posterior}.
	\INDSTATE Sample $vec\mathbb{G}_{(\mathcal{R}\times \mathcal{C})}^{(t)}$ according to \eqref{eq:post_g1}.
	\INDSTATE Get $\{\mathbf{U}^{(t)}_{(l)}\}_{l=1}^{L}, \{\mathbf{V}^{(t)}_{(m)}\}_{m=1}^{M}, \mathbb{G}^{(t)}$ according to \eqref{eq:vectorize}, \eqref{eq:matricize}.
	\INDSTATE Sample $\left( \sigma^2\right)^{(t)}$ according to \eqref{eq:sigma2_Posterior}.
	\INDSTATE Calculate $\mathbb{B}^{(t)}$ given by 
	\begin{equation*}
		\mathbb{B}^{(t)}=[\![\mathbb{G}^{(t)};\{\mathbf{U}_{(l)}^{(t)}\}_{l=1}^{L},\{\mathbf{V}_{(m)}^{(t)}\}_{m=1}^{M}]\!] \,.
	\end{equation*}
	\STATE Save $\left(\{\mathbf{U}^{(t)}_{(l)}\}_{l=1}^{L}, \{\mathbf{V}^{(t)}_{(m)}\}_{m=1}^{M}, \mathbb{G}^{(t)},  \mathbb{B}^{(t)}, \left(\sigma^2\right)^{(t)} \right)\,.$
	\ENDFOR				
\end{algorithmic}
\end{algorithm}

We summarize the full MCMC sampler in Algorithm \ref{algo:one}. For predictive inference, it is straightforward based on the posterior samples obtained from Algorithm \ref{algo:one}. Given new data $\mathbb{X}_{new}$ of $\tilde{N}$ samples, we can easily sample $\mathbb{Y}_{new}$ predictions  according to 
\begin{equation}
\label{eq:post_y_sample}
vec \mathbb{\hat{Y}}_{new} \ \sim \ N( vec\left(\langle \mathbb{X}_{new},\hat{\mathbb{B}} \rangle_L\right),\hat{\sigma}^{2}\mathbf{I}_{\tilde{N}Q}) \,,
\end{equation}
where $Q=\prod_{m=1}^{M}Q_m$, and $\hat{\mathbb{B}}$ is calculated by \eqref{eq:tucker_1} using post-burn-in samples of $\mathbb{G}, \{\mathbf{U}_{(l)}\}_{l = 1}^{L}, \{\mathbf{V}_{(m)}\}_{m = 1}^{M},$ and $\sigma^{2}$.

\section{Fast Computing Algorithm}
\label{sec:fast}		
In practice, the proposed MCMC sampler involves generating samples from high-dimensional conditional posterior distributions at each iteration, which can be  time-consuming. In this section, we propose an ultra-fast optimization-based computing algorithm as an alternative for posterior inference using the maximum a posteriori probability (MAP) estimators. 
Given the dimension of the core tensor, the MAP estimators of $\{\mathbf{U}_{(l)}\}_{l=1}^{L}, \{\mathbf{V}_{(m)}\}_{m=1}^{M}, \mathbb{G}$, and $\sigma^2$ can be computed in closed forms. Specifically, the MAP estimator of $\mathbf{U}_{(1)}$ is 
\begin{equation}
\label{eq:Fast_U}
vec\mathbf{U}_{(1)}^{MAP} = \left(\mathbb{C}_{(\mathcal{R}\times \Ccal)}^T\mathbb{C}_{(\mathcal{R}\times \Ccal)}+ \sigma^2 \mathbf{\Sigma}_U^{-1}\right)^{-1}\mathbb{C}_{(\mathcal{R}\times \Ccal)}^Tvec\mathbb{Y},
\end{equation}
when $\bm{\mu}_U=0$. 
And if we further set $\bm{\Sigma}_U$ to be an identity matrix, the result is exactly the solution to the ridge linear regression problem
$$\arg \min_{\mathbf{U}_{(1)}} \norm{\mathbb{Y}-\langle \mathbb{X,B} \rangle_L}^2_F-\lambda \norm{\mathbf{U}_{(1)}}^2_2,$$
with $\lambda=\sigma^2$ and $\mathbb{B}=[\![\mathbb{G};\mathbf{U}_{(1)},\cdots,\mathbf{U}_{(L)},\mathbf{V}_{(1)},\cdots,\mathbf{V}_{(M)}]\!]$. Similarly, assuming $\bm{\mu}_V = 0$, and $\bm{\mu}_G = 0$, the MAP estimators of $\mathbf{V}_{(1)}$ and the core tensor $\mathbb{G}$ are given below:
\begin{equation}
\label{eq:Fast_V}
\begin{split}
	{vec\mathbf{V}^T_{(1)}}^{MAP} & = \\
	& \left(\mathbf{I}_{Q_1} \otimes \left(\mathbb{D}^T_{(\mathcal{R}\times \Ccal)} \mathbb{D}_{(\mathcal{R}\times \Ccal)}\right) + \sigma^2\tilde{\mathbf{\Sigma}}_V^{-1}  \right)^{-1} \\
	& \times \left(\mathbf{I}_{Q_1}\otimes\mathbb{D}^T_{(\mathcal{R}\times \Ccal)}\right) vec\mathbf{\tilde{Y}} \,,
\end{split}
\end{equation}
and
\begin{equation}
\label{eq:Fast_G}
\begin{split}
	vec\mathbb{G}^{MAP}_{(\mathcal{R}\times \Ccal )}&= \\
	&\left( \left(\mathbf{V}^T\mathbf{V}\right)\otimes \left((\mathbb{X}_{(1)}\mathbf{U})^T
	(\mathbb{X}_{(1)}\mathbf{U})\right) + \sigma^2 \tilde{\bm{\Sigma}}_G^{-1} \right)^{-1} \\
	& \times \left(\left(\mathbf{V}^T\mathbf{V}\right)\otimes(\mathbb{X}_{(1)}\mathbf{U})^T \right)vec\dbtilde{\mathbf{Y}}  \,,
\end{split}
\end{equation}
with the same notations given in section \ref{sec:mcmc1}.
And the MAP estimator of $\sigma^2$ is given by
\begin{equation}
\label{eq:Fast_Sigma}
\left(\sigma^2\right)^{MAP} = \frac{\beta'}{\alpha' - 1}
\end{equation}
with $\alpha'=\alpha+\frac{NQ}{2}$,  $\beta'=\beta+\frac{\norm{\mathbb{Y}-\langle \mathbb{X,B}\rangle_L}^2_F}{2}$, and $Q=\prod_{m=1}^{M}Q_m$.
We remark that the unregularized least square results in \cite{lock2018tensor} is a special case of our MAP results above, when flat priors are given to $\mathbf{U}_{(l)}$'s and $\mathbf{V}_{(m)}$'s, and the core tensor is fixed to be a superdiagonal tensor.

By using MAP estimators of $\{\mathbf{U}_{(l)}\}_{l=1}^{L},\{\mathbf{V}_{(m)}\}_{m=1}^{M}$ and  $\mathbb{G}$ instead of generating samples from high dimensional posterior distributions, the problem of choosing the optimal dimension of the core tensor becomes a discrete optimization problem to find the tuple of parameters over an $(L+M)$-dimensional grid of parameters $\bm{\theta}:=(R_1,\cdots,R_L,S_1,\cdots,S_M)\in \bm{\Theta}$ that minimizes some loss function. In this work, we use the Bayesian information criterion (BIC) as the loss function. 

To solve the discrete optimization problem, we adopt simulated annealing (SA) algorithm \citep{kirkpatrick1983optimization}, which is a metaheuristic to approximate global optimum in a large search space. Starting from the current guess $\bm{\theta}^{(t)}$ at iteration $t$, we uniformly generate $\tilde{\bm{\theta}}$ from $O(\bm{\theta}^{(t)})$. 
The probability of accepting the new candidate $\tilde{\bm{\theta}}$ is $A(\tilde{\bm{\theta}},\bm{\theta}^{(t)})$ defined as follows:
\begin{equation}
\label{eq:SA_prob}
\begin{split}
	A(\tilde{\bm{\theta}},\bm{\theta}^{(t)}) &= \\
	&\begin{cases}
		1 & \text{if } BIC(\tilde{\bm{\theta}})<BIC(\bm{\theta}^{(t)}), \\
		\exp(\frac{BIC(\bm{\theta}^{(t)})-BIC(\tilde{\bm{\theta}})}{\zeta(t)}) & \text{otherwise},
	\end{cases} 
\end{split}
\end{equation}
where $\zeta(t)$, a function of the iteration $t$, is the usual \textit{temperature} parameter in standard SA algorithm. Two commonly-used choices for $\zeta(t)$ are $\zeta(t)=\gamma^t \zeta_0$ \citep{dosso1991magnetotelluric}, and $\zeta(t)=\frac{\zeta_0}{\log(1+t)}$  \citep{geman1984stochastic}, where $\zeta_0$ is the initial temperature. The detailed procedure of the fast computing algorithm is shown in Algorithm \ref{algo:three}. 

\begin{algorithm}[tbh]
\caption{Fast Computing Algorithm}
\label{algo:three}
\begin{algorithmic}[1]
	\renewcommand{\algorithmicrequire}{\textbf{Input:}}
	\renewcommand{\algorithmicensure}{\textbf{Output:}}
	\STATE Input data $\mathbb{X} \in \Rbb^{N\times P_1\times \cdots \times P_L}$, and $\mathbb{Y} \in \Rbb^{N \times Q_1\times \cdots \times Q_M}$.
	\STATE Initialize core tensor dimensions $\bm{\theta}^{(0)}$.
	\FOR{$t = 1,\cdots, T$}
	\STATE Calculate the temperature of current SA step: $\zeta(t)$.
	\STATE Generate $\tilde{\bm{\theta}}$ from the neighbour of $\bm{\theta}^{(t)}$.
	\FOR{$k = 1,\cdots, K$}
	\STATE Calculate $\{\tilde{\mathbf{U}}^{(k)}_{(l)}\}_{l=1}^{L}$ using \eqref{eq:Fast_U} and \eqref{eq:vectorize}.
	\STATE Calculate $\{\tilde{\mathbf{V}}^{(k)}_{(m)}\}_{m=1}^{M}$ using \eqref{eq:Fast_V} and \eqref{eq:vectorize}.
	\STATE Calculate
	$\tilde{\mathbb{G}}^{(k)}$ using \eqref{eq:Fast_G} and \eqref{eq:matricize}.
	\STATE Calculate $\left(\tilde{\sigma}^{2}\right)^{(k)}$ using \eqref{eq:Fast_Sigma}.
	\ENDFOR
	\STATE Given $\{\tilde{\mathbf{U}}^{(K)}_{(l)}\}_{l=1}^{L}, \{\tilde{\mathbf{V}}^{(K)}_{(m)}\}_{m=1}^{M}, \tilde{\mathbb{G}}^{(K)},\left(\tilde{\sigma}^{2}\right)^{(K)}$, and $\tilde{\bm{\theta}}$ calculate BIC.
	\STATE Calculate acceptance probability $A(\tilde{\bm{\theta}},\bm{\theta}^{(t)})$ given by \eqref{eq:SA_prob}.
	\STATE Generate $r\sim \mathrm{Unif}(0,1)$.
	\IF{$r < A(\tilde{\bm{\theta}},\bm{\theta}^{(t)})$}
	\STATE $\bm{\theta}^{(t+1)},\mathbb{G}^{(t+1)}, \left(\sigma^{2}\right)^{(t+1)}  = \tilde{\bm{\theta}}, \tilde{\mathbb{G}}^{(K)}, \left(\tilde{\sigma}^{2}\right)^{(K)} \,.$
	\STATE $\{\mathbf{U}_{(l)}^{(t+1)}\}_{l=1}^{L} = \{\tilde{\mathbf{U}}_{(l)}^{(K)}\}_{l=1}^{L}$.
	\STATE $\{\mathbf{V}_{(m)}^{(t+1)}\}_{m=1}^{M} = \{\tilde{\mathbf{V}}_{(m)}^{(K)}\}_{m=1}^{M}$.
	\ELSE
	\STATE $\bm{\theta}^{(t+1)},\mathbb{G}^{(t+1)}, \left(\sigma^{2}\right)^{(t+1)}  =\bm{\theta}^{(t)},\mathbb{G}^{(t)}, \left(\sigma^{2}\right)^{(t)}$.
	\STATE $\{\mathbf{U}_{(l)}^{(t+1)}\}_{l=1}^{L}, \{\mathbf{V}_{(m)}^{(t+1)}\}_{m=1}^{M} =   \{\mathbf{U}_{(l)}^{(t)}\}_{l=1}^{L}, \{\mathbf{V}_{(m)}^{(t)}\}_{m=1}^{M} $.
	\ENDIF
	\ENDFOR
	\STATE Calculate $\mathbb{B}^{(T)} = [\![\mathbb{G}^{(T)};\mathbf{U}_{(1)}^{(T)},\cdots,\mathbf{U}_{(L)}^{(T)},\mathbf{V}_{(1)}^{(T)},\cdots,\mathbf{V}_{(M)}^{(T)}]\!]$.
\end{algorithmic}
\end{algorithm}

As will be shown in numerical studies of section 6, the proposed Algorithm \ref{algo:three} performs well in terms of parameter estimation, and saves $>$90\% of the running time in most simulation cases. We also remark that we only obtain a point estimate for parameter estimation from Algorithm \ref{algo:three} since it is an optimization-based method. If one wants uncertainty quantification of parameters, we suggest to use the estimated optimal dimension of the core tensor from Algorithm \ref{algo:three}, and conduct the MCMC sampler introduced in section \ref{sec:mcmc1}, i.e., update model parameters given the estimated dimension of the core tensor. Such a treatment is able to not only save the computing time for updating the dimension of the core tensor in  Algorithm \ref{algo:one}, but also quantify parameter uncertainties.

\section{Simulation Studies}
\label{sec:simu}
We evaluate the proposed Bayesian tensor-on-tensor regression model and the computational algorithms through extensive simulation studies and comparisons to alternative methods. 

\subsection{Simulation setup}
Assume that the response tensor is a three-way array $\mathbb{Y} \in \Rbb^{N\times Q_1\times Q_2}$ and the predictor tensor is another three-way array $\mathbb{X} \in \Rbb^{N\times P_1 \times P_2}$.   The sample size is set to $N=100$. We set $P_1=16$, $P_2=12$, $Q_1=10$, and $Q_2=8$. 
We generate simulated data as follows:
\begin{itemize}
\item Generate $\mathbb{X}\in \Rbb^{N\times P_1\times P_2}$. 
\item Generate $\mathbf{U}_{(l)}$ and $\mathbf{V}_{(m)}$ for $l=1,\cdots,L$ and $m=1,\cdots,M$, each with independent $N(0,1)$ entries.
\item Generate $\mathbb{G}$ with $\theta^*$ dimensions, and independent $N(0,1)$ entries.
\item Generate $\mathbb{E} \in \Rbb^{N\times Q_1 \times Q_2}$ with $N(0,1)$ entries.
\item Calculate $\mathbb{Y}=\langle \mathbb{X,B}\rangle_L+c\mathbb{E}$, where 
\begin{equation*}
	\mathbb{B}=[\![\mathbb{G};\mathbf{U}_{(1)},\cdots,\mathbf{U}_{(L)},\mathbf{V}_{(1)},\cdots,\mathbf{V}_{(M)}]\!],
\end{equation*} and $c$ is a scaling parameter for defining the signal-to-noise (SNR) ratio such that 
$$\frac{\norm{\langle \mathbb{X,B}\rangle_L}_F^2}{c^2\norm{\mathbb{E}}_F^2}=\mathrm{SNR}.$$
\end{itemize}

We consider two different setups for generating the predictor tensor $\mathbb{X}$. In the first setup, all entries of $\mathbb{X}$ are independently and identically generated from a normal distribution $N(0, 1)$, called {\it uncorrelated setup}. In the second setup, we consider the entries for each observation of tensor $\mathbb{X}$ to have a correlated structure, which can be seen in some real-world applications such as spatial and temporal data \citep{lankao2008development}. We call it {\it correlated setup}.  Specifically, for each of the $N$ observations of $\mathbb{X}$, denoted by $\mathbf{X}_{(n)} \in \Rbb^{P_1\times P_2}$,   $n = 1,\cdots,N$, the correlation between the $(i,j)$ entry and $(k,l)$ entry of $\mathbf{X}_{(n)}$ is $e^{-r \sqrt{|i-k|^2+|j-l|^2}}$, where $i,k = 1,\cdots, P_1$, $j,l = 1,\cdots, P_2$, and $r > 0$.


In each of the setups, we consider two cases for the simulated true dimension of the core tensor $\mathbb{G}$: $\theta^*=(3,3,3,3)$ and  $(4,4,2,2)$, and two cases for SNR $=2$ and $5$, yielding a total of four cases.  We generate 50 replicated datasets for each case, and apply the proposed Bayesian tensor-on-tensor method using both the MCMC sampler (BayTensor MCMC) in Algorithm \ref{algo:one} and the fast computing Algorithm \ref{algo:three} (BayTensor Fast) for each dataset. 

To evaluate the performance of different methods, for each dataset in each case we generate 5 new datasets with $N_{new}=1000$ observations. The new observations are given by 
$$\mathbb{Y}_{new}=\langle \mathbb{X}_{new},\mathbb{B}\rangle_L+c\mathbb{E}_{new},$$ with $\mathbb{X}_{new}$ and $\mathbb{E}_{new}$ generated in the same way as $\mathbb{X}$ and $\mathbb{E}$. We then calculate the relative prediction error (RPE), defined as the average prediction error for the 5 new datasets: 
$$\mathbf{RPE}=\frac{1}{5}\sum_{i=1}^5\frac{\norm{\mathbb{Y}^{(i)}_{new}- \hat{\mathbb{Y}}^{(i)}_{new}}_F^2}{\norm{\mathbb{Y}^{(i)}_{new}}_F^2} \,,$$
where $\hat{\mathbb{Y}}^{(i)}_{new}$ is the predicted value for the dataset $i$, $i=1, \dots, 5.$

For comparison, we consider two alternative methods: 
\begin{itemize}
\item The tensor-on-tensor regression method based on the CP decomposition from \cite{lock2018tensor}, denoted as the CP method. As the CP method needs to pre-define or estimate the rank $R$, we run their algorithm with different $R$'s and report the result under the optimal $R$ value, defined as the one yielding the smallest RPE. 
\item Multivariate linear regression method after turning tensors to vectors to solve  $\mathbb{B}_{(\mathcal{P}\times \mathcal{Q})}$ in equation \eqref{eq:tensor_reg2}, denoted as the OLS method.
\end{itemize}


\subsection{Simulation results: uncorrelated setup}
We apply the proposed BayTensor MCMC, BayTensor Fast, the CP method, and the OLS method to datasets under the uncorrelated setup. For the CP method, we run the algorithm with $R = 1,\cdots,6$. The reason why we choose 6 as the upper bound is that when $R = 6$, the total number of parameters in the CP method is 276 which is larger than the true number of parameters 212 for the $\bm{\theta}^* = (4,4,2,2)$ case and 219 for the $\bm{\theta}^* = (3,3,3,3)$ case. The CP method with $R = 6$ yields the smallest RPE under both cases in all repeated simulations. Full results from the CP method with different $R$ values are shown in Appendix \ref{sec:CP_rank}. 

Table \ref{tbl:one} reports the means and standard deviations (sd) of RPEs averaging over 50 replicated experiments for each of the four cases under the four methods. 
No matter when the information in $\mathbb{X}$ and $\mathbb{Y}$ is balanced (i.e., $\theta^*=(3,3,3,3)$) or skewed to some modes (i.e.,  $\theta^*=(4,4,2,2)$), BayTensor MCMC and BayTensor Fast always yield smaller prediction RPEs than the CP method, while the OLS method performs the worst. For example, when $\theta^*=(4,4,2,2)$ with SNR $=2$, BayTensor MCMC and BayTensor Fast have comparable results with mean RPEs being 0.350  and 0.357 respectively, while the CP method has a slightly larger mean RPE of 0.377 and the OLS method has the largest mean RPE of 1.016. The superior performance of the Bayesian tensor-on-tensor regression method over the CP method comes from the flexibility of Tucker decomposition that permits different orders along different modes of the core tensor. From Table \ref{tbl:one}, we can also see that a larger signal to noise ratio leads to lower RPEs. In particular, when SNR $=5$, all methods have lower  RPEs compared to the results from cases where SNR $=2$. 


As the proposed BayTensor MCMC method and BayTensor Fast method can simultaneously estimate the dimension of the core tensor and other model parameters, we next report the empirical probabilities that the true dimension of the core tensor can be recovered by them in 50 replicated experiments for all 4 cases, as shown in Table \ref{tbl:two}. BayTensor MCMC has higher recovering rates than BayTensor Fast in all cases. And we can better recover the core tensor dimension with a higher SNR. Another observation is that both BayTensor MCMC and BayTensor Fast have higher recovering rates when $\theta^*=(3,3,3,3)$ compared with the cases where $\theta^*=(4,4,2,2)$. 

We also present the average numbers of parameters required by BayTensor MCMC and BayTensor Fast in 50 replicated experiments in Table \ref{tbl:two}. BayTensor MCMC and BayTensor Fast usually require smaller numbers of parameters than the CP method. For example, when $\theta^*=(4,4,2,2)$ with SNR $=2$, BayTensor MCMC requires 207 parameters on average, and BayTensor Fast requires 196 parameters on average. In contrast, the CP method with $R = 6$ has 276 parameters. When $\theta^*=(3,3,3,3)$ with SNR $=2$, the average numbers of parameters are 218, 193, and 276 for BayTensor MCMC, BayTensor Fast, and the CP method respectively. We can see that both BayTensor MCMC and BayTensor Fast yield higher dimension recovery rates than the CP method with smaller numbers of parameters.

\begin{table*}[!htpb]
\centering
\caption{Mean RPE with SD on uncorrelated data.}

\resizebox{\textwidth}{!}{\begin{tabular}{c|cccc}
		\hline 
		RPE (SD) & BayTensor MCMC & BayTensor Fast & CP Method & OLS Method \\ 
		\hline 
		$\theta^*=(4,4,2,2)$, SNR=2 & \textbf{0.350 (0.013)} & 0.357 (0.026) & 0.377 (0.014) & 1.016 (0.034)  \\ 
		\hline 
		$\theta^*=(3,3,3,3)$, SNR=2 & \textbf{0.343 (0.004)} & 0.351 (0.015) & 0.391 (0.016) & 1.023 (0.031)  \\ 
		\hline 
		$\theta^*=(4,4,2,2)$, SNR=5 & \textbf{0.173 (0.001)}& 0.185 (0.034) &0.209 (0.046) & 0.751 (0.030)  \\ 
		\hline 
		$\theta^*=(3,3,3,3)$, SNR=5 & \textbf{0.172 (0.001)} & 0.181 (0.037) & 0.222 (0.016) & 0.748 (0.023)  \\ 
		\hline 
\end{tabular}}
\label{tbl:one}
\end{table*}

\begin{figure}[H]
\centering
\begin{subfigure}{0.9\textwidth}
	\includegraphics[width=\textwidth]{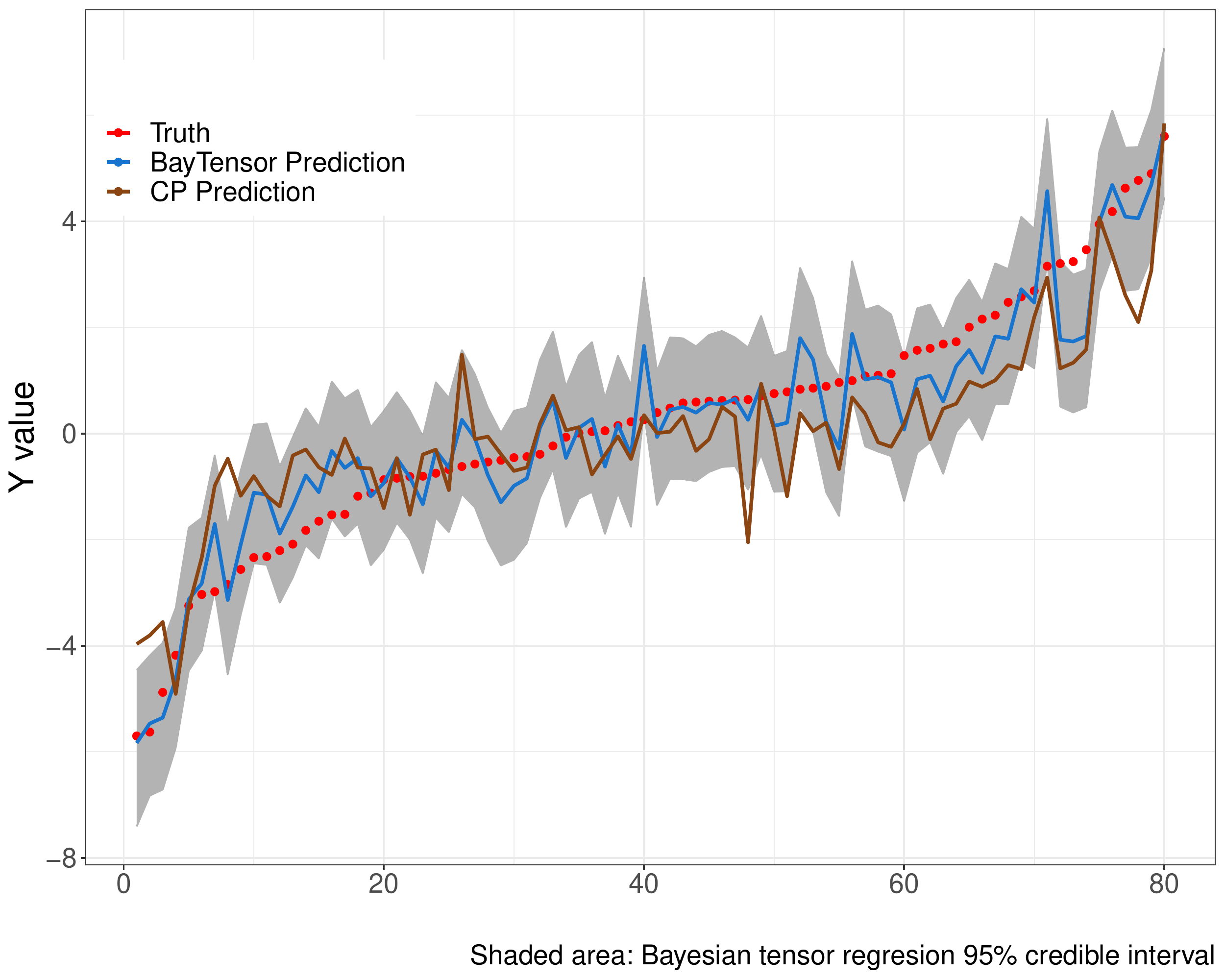}
	\caption{}
\end{subfigure}
\begin{subfigure}{0.9\textwidth}
	\includegraphics[width=\textwidth]{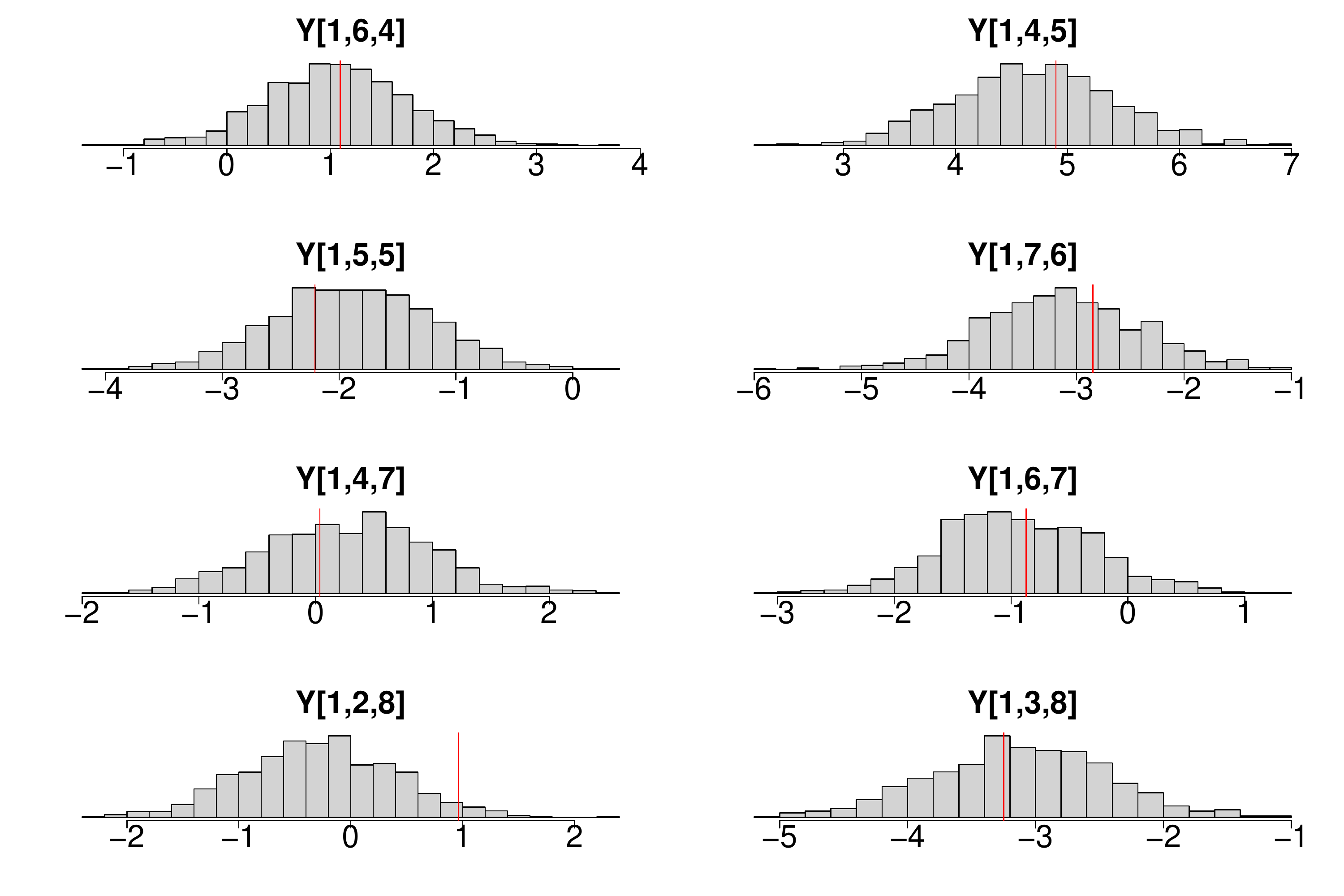}
	\caption{}
\end{subfigure}
\caption{Uncertainty quantification for a randomly selected case where $\bm{\theta}^* = (3,3,3,3)$ and SNR $=5$. (a): An example of prediction with estimation uncertainty. Elements of $\mathbb{Y}_{[1::]}$ are sorted for visualization. (b): Empirical predictive distributions from BayTensor MCMC for randomly selected elements from $\mathbb{Y}_{new}$ with the simulation truths shown as red lines.}
\label{fig:UQ_1}
\end{figure}

\begin{table*}[!htpb]
\centering
\caption{Core tensor dimension recovery and number of model parameters on uncorrelated data.}
\resizebox{\textwidth}{!}{\begin{tabular}{c|cc|cc}
		\hline 
		& \multicolumn{2}{c|}{BayTensor MCMC}  & \multicolumn{2}{c}{BayTensor Fast} \\ 
		\hline
		& Dimension Recovery & \# Parameters (SD) & Dimension Recovery & \# Parameters (SD) \\
		\hline 
		$\theta^*=(4,4,2,2)$, SNR=2 & $80\%$ & 207 (15) & $46\%$ & 189 (14) \\ 
		\hline 
		$\theta^*=(3,3,3,3)$, SNR=2 & $98\%$ & 218 (6) & $54\%$ & 193 (18)  \\ 
		\hline 
		$\theta^*=(4,4,2,2)$, SNR=5 & $96\%$ & 215 (14) & $64\%$ & 209 (16)  \\ 
		\hline 
		$\theta^*=(3,3,3,3)$, SNR=5 & $100\%$ & 219 (0)  & $76\%$ & 215 (16)  \\ 
		\hline 
\end{tabular}}
\label{tbl:two}
\end{table*}

To quantify the prediction uncertainty, for each test dataset under each simulation setup, we collect 1000 post-burn-in samples from BayTensor MCMC and generate the empirical posterior predictive distribution for each element of $\mathbb{Y}_{new}$. The symmetric credible intervals are then calculated. For cases with $\bm{\theta}^* = (4,4,2,2)$, the empirical coverage rates for $95\%$ credible interval are 0.944 (0.013) and 0.945 (0.012) for SNR $=2$ and SNR $=5$ respectively. And for cases with $\bm{\theta}^* = (3,3,3,3)$, the empirical coverage rates for $95\%$ credible interval are 0.950 (0.011) and  0.950 (0.011) for SNR = $2$ and SNR = $5$ respectively.
Figure \ref{fig:UQ_1}(a) plots the predictive posterior estimations with 95\% credible intervals for a randomly-selected sample of $\mathbb{Y}_{new}$ under a randomly-selected test case when $\bm{\theta}^* = (3,3,3,3)$ and SNR $=5$. Figure \ref{fig:UQ_1}(b) plots the empirical predictive distributions from BayTensor MCMC for 8 randomly selected elements from the same sample of $\mathbb{Y}_{new}$ shown in Figure \ref{fig:UQ_1}(a).


\subsection{Simulation results: correlated setup}

\begin{table*}[!htpb]
\centering
\caption{Mean RPE with SD on correlated data.}
\resizebox{\textwidth}{!}{\begin{tabular}{c|cccc}
		\hline 
		RPE (SD) & BayTensor MCMC & BayTensor Fast & CP Method & OLS Method \\ 
		\hline 
		$\theta^*=(4,4,2,2)$, SNR=2 & \textbf{0.344 (0.004)} & 0.350 (0.006) & 0.370(0.011) & 0.886 (0.086)  \\ 
		\hline 
		$\theta^*=(3,3,3,3)$, SNR=2 & \textbf{0.344 (0.004)}& 0.348 (0.006) &0.369(0.007) &  0.893 (0.065) \\ 
		\hline 
		$\theta^*=(4,4,2,2)$, SNR=5 & \textbf{0.173 (0.002)} & 0.175 (0.003)  & 0.188(0.006) &  0.492 (0.086) \\ 
		\hline 
		$\theta^*=(3,3,3,3)$, SNR=5 & \textbf{0.173 (0.008)} & 0.173 (0.003) & 0.187(0.006) & 0.497 (0.059)  \\ 
		\hline 
\end{tabular}}
\label{tbl:three}
\end{table*}

We then apply the proposed BayTensor MCMC method, the BayTensor Fast method, the CP method, and the OLS method to datasets under the correlated setup. For the CP method, we run the algorithm with $R = 1,\cdots,6$. The CP method with $R = 6$ yields the smallest RPE under all cases except for the case of $\bm{\theta}^* = (4,4,2,2)$ and SNR $=2$ where the optimal $R = 5$. Full results from the CP method with different $R$ values are shown in Appendix \ref{sec:CP_rank}. For each replicated dataset, we compute the  RPEs under all four methods. The 
means and standard deviations (sd) of RPEs averaging over the 50 replicated datasets for all 4 cases are presented in Table \ref{tbl:three}.

The prediction RPE results under the correlated setup are analogous to the results under  the uncorrelated setup. To summarize, the RPE results from Baytensor MCMC  and BayTensor Fast are comparable in all cases. And both of them have better RPEs than the CP method. The OLS performs the worst in terms of yielding the highest RPEs in all cases. And when the signal to noise ratio increases from 2 to 5, the prediction accuracies are improved under all methods. 

Table \ref{tbl:four} shows the empirical probabilities of dimension recovery and the average numbers of parameters required by BayTensor MCMC and BayTensor Fast in 50 replicated experiments. In all 4 cases, BayTensor MCMC and BayTensor Fast both require a smaller number of parameters than the CP method.
And in terms of recovering the dimension of the  core tensor, BayTensor MCMC has higher empirical probabilities of recovering the true  dimension than BayTensor Fast in all cases. And the recovering probabilities in the cases where $\bm{\theta}^* = (3,3,3,3)$ are higher than those in cases where $\bm{\theta}^* = (4,4,2,2)$ for both methods. We also observe that the recovering probabilities under the correlated setup are smaller than those under the uncorrelated setup for both the BayTensor MCMC and BayTensor Fast.

\begin{table*}[!htpb]
\centering
\caption{Core tensor dimension recovery and number of model parameters on correlated data.}
\resizebox{\textwidth}{!}{\begin{tabular}{c|cc|cc}
		\hline 
		& \multicolumn{2}{c|}{BayTensor MCMC}  & \multicolumn{2}{c}{BayTensor Fast} \\ 
		\hline
		& Dimension Recovery & \# Parameters (SD) & Dimension Recovery & \# Parameters (SD) \\
		\hline 
		$\theta^*=(4,4,2,2)$, SNR=2 & $72\%$ & 207 (17) & $36\%$ & 196 (16) \\ 
		\hline 
		$\theta^*=(3,3,3,3)$, SNR=2 & $86\%$ & 214 (14) & $32\%$ & 164 (20)  \\ 
		\hline 
		$\theta^*=(4,4,2,2)$, SNR=5 & $96\%$ & 212 (7) & $44\%$ & 188 (15)  \\ 
		\hline 
		$\theta^*=(3,3,3,3)$, SNR=5 & $96\%$ & 218 (6)  & $56\%$ & 196 (15)  \\ 
		\hline 
\end{tabular}}
\label{tbl:four}
\end{table*}


\section{Real data analyses}
\label{sec:real}
We demonstrate the usefulness of the proposed Bayesian tensor-on-tensor regression approach by applying it to two real-world datasets: labeled faces in the wild database (\cite{LFWTech}) and multi-person motion (UMPM) benchmark \citep{van2011umpm}. 

\subsection{Facial image data}
We apply the proposed Bayesian tensor-on-tensor regression approach to predict different attributes of facial images, such as smiling and nose size, using the labeled faces in the wild database (\cite{LFWTech}). The database collects more than 13,000 face images, each of which has been labeled with the name of the person pictured, often a celebrity. There can be multiple images for one person. 

We use the attribute classifiers developed in \cite{kumar2009attribute} for attributes $\mathbb{Y}$, resulting in a total of 73 describable attributes. These attributes can be categorized into characteristics that describe a person, an  expression, or an accessory. The attributes are all given as continuous variables, with a higher value denoting a more obvious characteristic. The proposed approach is applied to predict the 73 attributes from a given facial image $\mathbb{X}$. In this work, we use the frontalized version of facial images \citep{hassner2015effective}, which show only forward-facing faces obtained by rotating, scaling, and cutting original facial images. The frontalized images are highly aligned, allowing for appearances to be easily compared across faces. Each frontalized image contains $90\times 90$ pixels, with each pixel giving color intensities for red, green, and blue, resulting in a $90\times 90 \times 3$ tensor for each image. We randomly sample $1000$ images. Thus the predictor tensor $\mathbb{X}$ is of dimensions $1000\times 90 \times 90 \times 3$, and the response tensor $\mathbb{Y}$ is of dimensions $1000 \times 73$. We center the tensor for each image by subtracting the mean of tensors for all images. Another randomly-sampled $1000$ images are used as a validation set, in other words, $\mathbb{X}_{new}$ is of dimensions $1000 \times 90 \times 90 \times 3$, and $\mathbb{Y}_{new}$ is of dimensions $1000 \times 73$. 



We apply the BayTensor MCMC in Algorithm \ref{algo:one} and BayTensor Fast in Algorithm \ref{algo:three} to the dataset and conduct inference. For comparison, we also apply the CP method proposed by \cite{lock2018tensor} to the same dataset. For the CP method, we choose $R = 15$ and $\lambda = 10^5$ since these values yielded the best prediction performance, as reported in \cite{lock2018tensor}. BayTensor MCMC estimates the dimension of the core tensor based on the posterior mode to be $(5,2,3,5)$ with a total of $1154$ parameters, resulting in an RPE of $0.375$. And BayTensor Fast yields an RPE of 0.446. In contrast, the CP method has 3840 parameters and results in a higher RPE, 0.477. To summarize, the proposed Bayesian tensor-on-tensor regression approach is able to reduce predictive errors with a smaller number of parameters due to the flexibility of the Tucker decomposition that permits different orders along different modes.

\begin{figure}[H]
\centering
\includegraphics[width=0.9\textwidth]{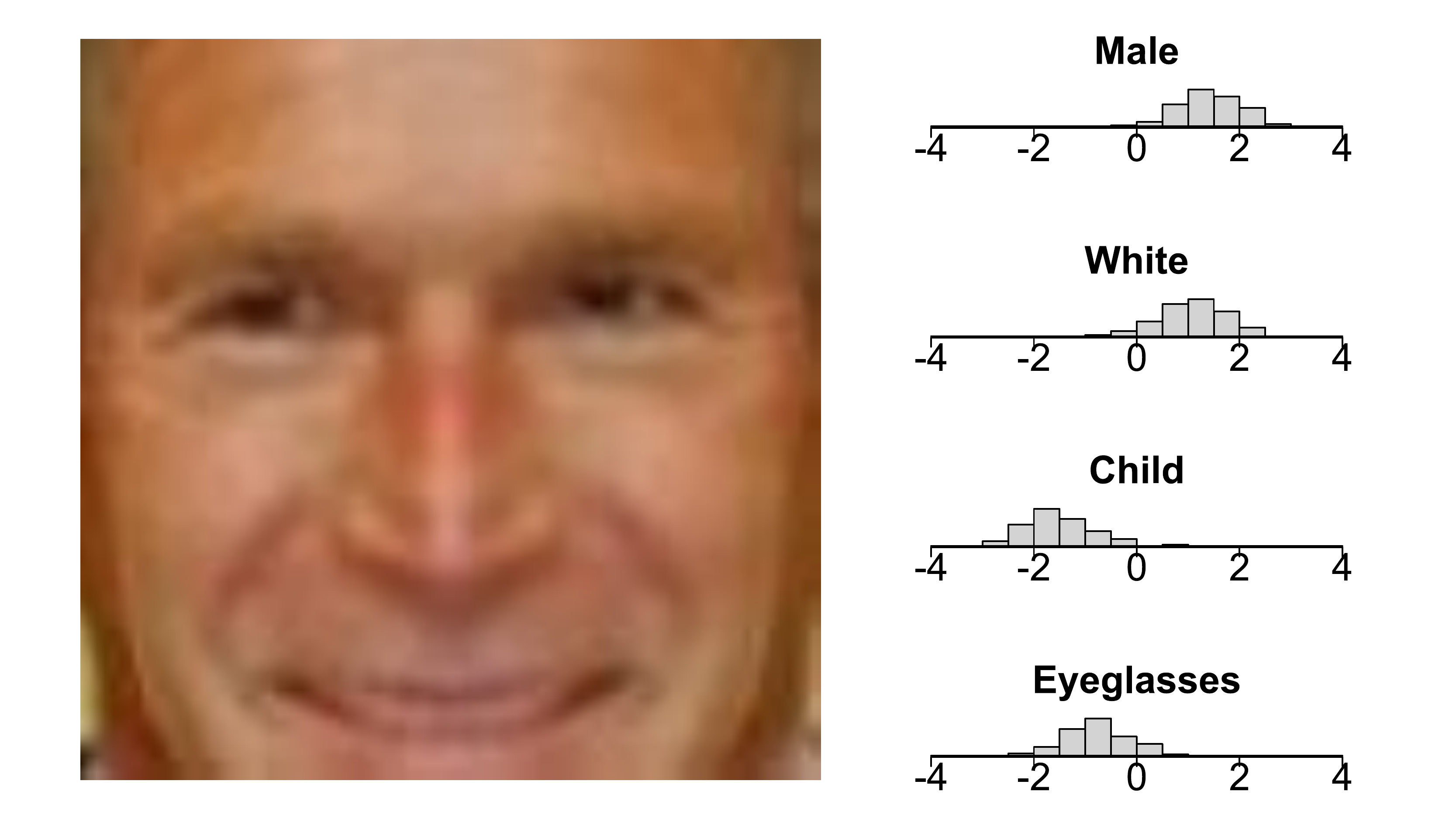}
\caption{An example of a test image and the corresponding posterior predictive values for four selected characteristics.}
\label{fig:face}
\end{figure}

Next we report the prediction uncertainty, which is a natural byproduct of the proposed Bayesian framework. We collect $500$ post-burn-in MCMC samples and compute the posterior predictive distribution along with credible intervals for each of the $73$ attributes for each image. 
The empirical coverage rate for $95\%$ credible interval is $0.930$, and for $90\%$ credible interval is $0.889$. 
Figure \ref{fig:face} shows an example of a test image, and plots the corresponding posterior predictive values for four randomly-selected characteristics.

\subsection{Utrecht Multi-Person Motion (UMPM) data}
We then apply the proposed Bayesian tensor-on-tensor regression model to the multi-person motion (UMPM) benchmark \citep{van2011umpm} that contains temporally synchronized video sequences from multiple viewpoints and human motion capture
data. 
Each video is of length 30 to 60 seconds, with resolution of $644 \times 484 $ pixels at 50 fps (frames per second). Motion capture (MoCap) data contain 3D positions of 37 markers at 100 fps for each subject of interest.

To evaluate the performance of our model on 3D motion data, we consider two scenarios, namely `chair' and `triangle' from the UMPM dataset. In the scenario `chair', we have a sequence of $2570$ sample images in which the subject of interest starts walking in a circle, finds the chair, sits on the chair and stands up multiple times with different postures. In the scenario `triangle', we have a sequence of $2471$ sample images in which the subject walks by following the path of a triangle within a circular area.
In our data analysis, the input data was a grayscale image sequence from the front camera with resolution downsized to $32\times 24$ pixels, forming a 3-order predictor tensor (i.e. \textit{frames} $\times$ \textit{width} $\times$ \textit{height}). The response data containing 3D positions of 37 markers is first downsampled to 50fps to match the video, and is then presented as a 3-order tensor (i.e. \textit{samples} $\times$ \textit{3D position} $\times$ \textit{markers}). 
For each scenario, we run 10 repeated experiments, and randomly sample 200 images from the sequence in each experiment to form the predictor tensor data $\mathbb{X}$ of dimensions $200 \times 32 \times 24$ and response tensor data $\mathbb{Y}$ of dimensions $200 \times 3 \times 37$. The remaining images are used as testing data $\mathbb{X}_{new}$ and $\mathbb{Y}_{new}$. 

We apply BayTensor MCMC, BayTensor Fast, and the CP method to the datasets. For scenario `chair', the mean prediction RPE of BayTensor MCMC is 0.176 (sd: 0.009) while that of BayTensor Fast and the CP method are 0.206 (sd: 0.053) and 0.254 (sd: 0.064) respectively. The mean number of parameters for BayTensor MCMC is 752 (sd: 58), for BayTensor Fast, the number is 798 (sd: 132), and for the CP method, with $R= 10$, the number of parameters is 960.
For scenario `triangle', BayTensor MCMC yields a mean RPE of 0.243 (sd: 0.049)  with the mean number of parameters being 685 (sd: $127)$. BayTensor Fast, with the mean number of parameters being 798 (sd: 132), has a slightly larger RPE of 0.252 (sd: 0.028). In contrast, the CP method, with $R = 10$, has 960 parameters and results in a worse RPE, 0.346 (sd: 0.049).  To summarize, with a smaller number of estimating parameters, the proposed Bayesian tensor-on-tensor regression model is able to yield smaller predictive errors compared to the CP method. And this improvement comes from the flexibility of Tucker decomposition.



\begin{figure}[H]
\centering
\includegraphics[width=0.9\textwidth]{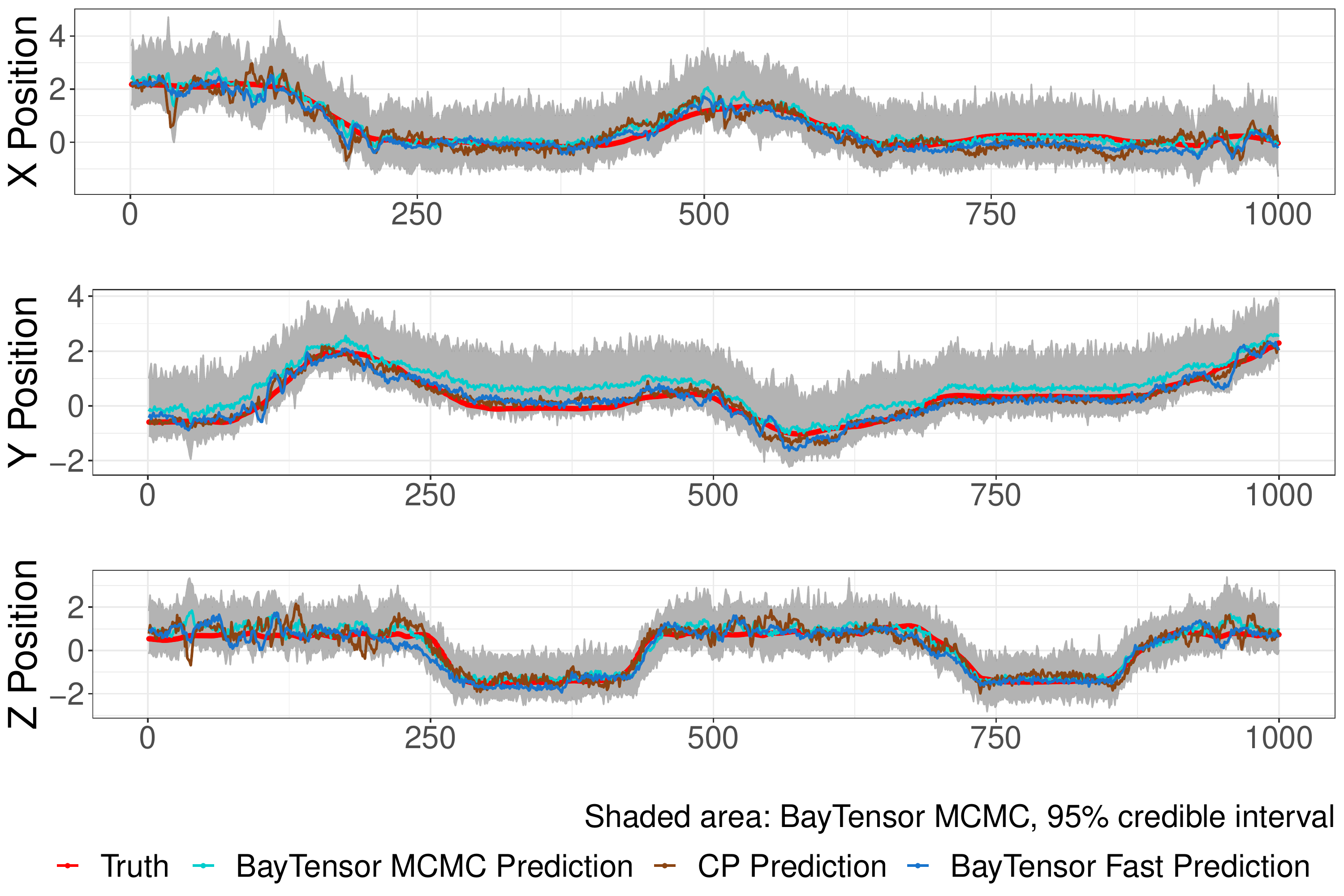}
\caption{Predictions of x,y,z-position of marker 3 by Bayesian tensor-on-tensor regression method with $95\%$ credible intervals, the fast computing method, and the CP method. The underlying truths are shown in red.}
\label{fig:motion}
\end{figure}

To demonstrate the predictive uncertainty, we generate 500 post-burn-in posterior samples, produce the empirical posterior predictive distribution of 3D position for each of the 37 markers, and calculate the symmetric credible intervals. The empirical coverage rate for 95$\%$ credible interval is $0.957$ for scenario `chair', and $0.931$ for scenario `triangle'.  Figure \ref{fig:motion} shows an example of estimation results for 3D position of the $\#$3 marker by BayTensor MCMC method with $95\%$ credible intervals, BayTensor Fast, and the CP method.

\section{Discussion}
\label{sec:dis}
We developed a Bayesian tensor-on-tensor regression model to predict one tensor response from another tensor predictor, building upon the flexible Tucker decomposition of the coefficient tensor so that the response tensor and predictor tensor can have different dimensions in the core tensor. For posterior inference, we proposed an efficient MCMC algorithm to simultaneously estimate the model dimension and model parameters. 
In addition, we developed an ultra-fast computing algorithm wherein the MAP estimators of model parameters are computed given the model dimension, and then an SA algorithm is used to find the optimal model dimension.
Both simulation studies and real data analyses show that the performance of our Bayesian tensor-on-tensor regression model benefits from the flexibility in the core tensor structure compared to alternative methods. And the fast computing algorithm yielded comparable prediction results to the MCMC sampler, meanwhile saved a significant amount of computing time. To our best knowledge, this work represents the first effort in literature to simultaneously estimate the model dimension and parameters in tensor-on-tensor regression setup under a  fully Bayesian framework. 

There are several interesting future directions. 
First, instead of assigning normal priors to factor matrices, the proposed framework can easily incorporate other priors, such as sparsity-inducing priors \citep{guhaniyogi2017bayesian, miranda2018tprm}, for applications where the sparsity is desired. Second, this work only considers one tensor predictor. Some applications may require to predict one tensor response from multiple tensor predictors or mixed-type predictors including tensors, matrices, and vectors. We will extend the proposed model to handle more flexible predictors. Finally, the computational cost of the proposed MCMC sampler comes as a price for prediction accuracy and uncertainty quantification compared to the proposed fast computing algorithm. To improve the efficiency of posterior computation, we plan to take advantage of certain advanced MCMC techniques, e.g., stochastic variational inference \citep{hoffman2013stochastic} and pseudo-marginal Metropolis-Hastings algorithms \citep{andrieu2009pseudo}.


\appendix
\section*{Appendix}
\label{appendix}
\subsection*{A brief review of matrix Kronecker product}
\label{sec:kro}
The \textit{Kronecker product} is a matrix operation that is important in showing posterior distribution of parameters in this paper, and we will briefly review it here. 

The Kronecker product of matrices $\mathbf{A}\in \Rbb^{I\times J}$, and $\mathbf{B}\in \Rbb^{K\times L}$ is denoted by $\mathbf{A}\otimes \mathbf{B}$. And the result is of size $(IK)\times (JL)$ defined by 
$$\mathbf{A}\otimes \mathbf{B}=\left[\begin{matrix}
a_{11}\mathbf{B} & a_{12}\mathbf{B} & \cdots & a_{1J}\mathbf{B} \\
a_{21}\mathbf{B} & a_{22}\mathbf{B} & \cdots & a_{2J}\mathbf{B} \\
\vdots & \vdots & \ddots & \vdots \\
a_{I1}\mathbf{B} & a_{I2}\mathbf{B} & \cdots & a_{IJ}\mathbf{B}
\end{matrix}\right].$$
Some of the properties of Kronecker product are proved useful for this paper. See the detailed proofs of these properties in \cite{kolda2006multilinear}
$$(\mathbf{A}\otimes \mathbf{B})(\mathbf{C}\otimes \mathbf{D})=\mathbf{AC}\otimes \mathbf{BD}.$$
Let $\mathbb{X}\in \Rbb^{I_1\times I_2 \times \cdots \times I_N}$, and $\Ncal =\{1,\cdots,N\}$. Let $\mathbf{A}_{(n)}\in \Rbb^{I_n\times J_n}$ be a sequence matrices for all $n\in \Ncal$.  Let the ordered sets $\mathcal{R}=\{r_1,\cdots,r_L\}$ and $\mathcal{C}=\{c_1,\cdots,c_M\}$ be a partition of $\Ncal$, then if $$\mathbb{X}=\mathbb{Y}\times_1 \mathbf{A}_{(1)}\times_2 \mathbf{A}_{(2)}\cdots \times_N \mathbf{A}_{(N)} $$ we have 
\begin{align*}
\mathbb{X}_{(\mathcal{R}\times \mathcal{C})}&=\left(\mathbf{A}_{(r_L)}\otimes \cdots \otimes \mathbf{A}_{(r_1)}\right)\mathbb{Y}_{(\mathcal{R}\times \mathcal{C})}\\
&\times \left(\mathbf{A}_{(c_M)}\otimes \cdots \otimes \mathbf{A}_{(c_1)}\right)^T.
\end{align*}
Consequently, if $\mathbf{A}_{(n)}\in \Rbb^{I_n\times J_n}$ for all $n\in \Ncal$, then for any specific $n\in \Ncal$ if we have $$\mathbb{X}=\mathbb{Y}\times_1 \mathbf{A}_{(1)}\times_2 \mathbf{A}_{(2)}\cdots \times_N \mathbf{A}_{(N)} ,$$ and then 
\begin{equation}
\label{eq:tucker_xn}
\begin{split}
	\mathbb{X}_{(n)}&=\mathbf{A}_{(n)}\mathbb{Y}_{(n)}\\
	& \times \left(\mathbf{A}_{(N)}\otimes \cdots \otimes \mathbf{A}_{(n+1)}\otimes \mathbf{A}_{(n-1)}\otimes \cdots \otimes \mathbf{A}_{(1)}\right)^T.
\end{split}
\end{equation}

\subsection*{Proof for equation \eqref{eq:post_u1}}
\label{sec:post_u1}
By properties of n-mode product of tensor and Tucker decomposition, we have
$$\mathbb{B}=\mathbb{G}\times_2 \mathbf{U}_{(2)} \cdots \times_L \mathbf{U}_{(L)} \times_{L+1} \mathbf{V}_{(1)} \cdots \times_{L+M} \mathbf{V}_{(M)} \times_1 \mathbf{U}_{(1)}.$$ 
Let $\mathbb{B}_{(-)}$ denote $\mathbb{G}\times_2 \mathbf{U}_{(2)} \cdots \times_L \mathbf{U}_{(L)} \times_{L+1} \mathbf{V}_{(1)} \cdots \times_{L+M} \mathbf{V}_{(M)}$, then $\mathbb{B}_{(-)}\in \Rbb^{R_1\times P_2\times\cdots \times P_L \times Q_1 \times \cdots \times Q_M}$, and
$$\mathbb{B}=\mathbb{B}_{(-)}\times_1 \mathbf{U}_{(1)}.$$
Therefore, 
\begin{align*}
&\mathbb{B}_{[p_1,\cdots,p_L,q_1,\cdots,q_M]} =\\
& \sum_{r_1=1}^{R_1}\left((\mathbb{B}_{(-)})_{[r_1,p_2,\cdots,p_L,q_1,\cdots,q_M]}\mathbf{U}_{(1)r_1p_1} \right).
\end{align*}
And \begin{align}
&\mathbb{\hat{Y}}_{[n,q_1,\cdots,q_M]}=\left(\langle \mathbb{X,B} \rangle_L \right)_{[n,q_1,\cdots,q_M]} \nonumber \\
&=\sum_{p_1=1}^{P_1} \cdots \sum_{p_L=1}^{P_L}\mathbb{B}_{[p_1,\cdots,p_L,q_1,\cdots,q_M]}\mathbb{X}_{[n,p_1,\cdots,p_L]} \nonumber \\
&=\sum_{p_1=1}^{P_1} \cdots \sum_{p_L=1}^{P_L}\sum_{r_1=1}^{R_1} \left( (\mathbb{B}_{(-)})_{[r_1,p_2,\cdots,p_L,q_1,\cdots,q_M]} \right. \nonumber \\
& \quad \quad \times \left. \mathbf{U}_{(1)r_1p_1}\mathbb{X}_{[n,p_1,\cdots,p_L]}\right) \nonumber \\
&=\sum_{r_1=1}^{R_1}\sum_{p_1=1}^{P_1} \nonumber \\
& \left(\sum_{p_2=1}^{P_2} \cdots \sum_{p_L=1}^{P_L}(\mathbb{B}_{(-)})_{[r_1,p_2,\cdots,p_L,q_1,\cdots,q_M]}\mathbb{X}_{[n,p_1,\cdots,p_L]}\right) \mathbf{U}_{(1)r_1p_1} \nonumber \\
&=\sum_{r_1=1}^{R_1}\sum_{p_1=1}^{P_1} \left(\langle \mathbb{B}_{(-)},\mathbb{X}\rangle_{P_2,\cdots,P_N} \right)_{[n,p_1,r_1,q_1,\cdots,q_M]} \mathbf{U}_{(1)r_1p_1}. \nonumber
\end{align}
And further, 
\begin{equation*}
vec\mathbb{\hat{Y}}= \mathbb{C}_{(\mathcal{R}\times \Ccal)} \times \ vec \mathbf{U}_{(1)}.
\end{equation*}

If we denote the contracted product of $\langle \mathbb{B}_{(-)},\mathbb{X}\rangle_{P_2,\cdots,P_N}$ by a new tensor called $\mathbb{C}$, then tensor 	$\mathbb{C}\in \Rbb^{R_1\times N \times P_1 \times Q_1\times \cdots Q_M}$. And matricize tensor $\mathbb{C}$ to $\mathbb{C}_{(\mathcal{R}\times \Ccal)} \in \Rbb^{N\prod_{m=1}^{M}Q_m \times R_1P_1}$, where $\mathcal{R}=\{N,Q_1,\cdots,Q_M\}$, and $\mathcal{C}=\{R_1,P_1\}$. We have 
\begin{equation*}
vec\mathbb{Y}= \mathbb{C}_{(\mathcal{R}\times \Ccal)} \times \ vec \mathbf{U}_{(1)}+vec\mathbb{E}.
\end{equation*}

\subsection*{Proof for equation \eqref{eq:post_v1}}
\label{sec:post_v1}
We denote tensor $\mathbb{D}_{(-)}=\mathbb{G}\times_1 \mathbf{U}_{(1)} \cdots \times_L \mathbf{U}_{(L)} \times_{L+2} \mathbf{V}_{(2)} \cdots \times_{L+M} \mathbf{V}_{(M)}$. And then the contracted product of tensor $\mathbb{D}_{(-)}$ and $\mathbb{X}$ is a new tensor denoted as $\mathbb{D}$. Then 
\begin{equation*}
\mathbb{B}=\mathbb{D}_{(-)}\times_{L+1} \mathbf{V}_{(1)} \,.
\end{equation*}
Then 	\begin{equation*}
\mathbb{B}_{[p_1,\cdots,p_L,q_1,\cdots,q_M]}=\sum_{s_1=1}^{S_1}{\mathbb{D}_{(-)}}_{[p_1,\cdots,p_L,s_1,q_2,\cdots,q_M]}\mathbf{V}_{(1)s_1 q_1} \,,
\end{equation*}
and
\begin{equation}
\label{eq:post_v2}
\begin{split}
	&\mathbb{\hat{Y}}_{[n,q_1,\cdots,q_M]}=\left(\langle \mathbb{X,B} \rangle_L \right)_{[n,q_1,\cdots,q_M]} \\
	&=\sum_{p_1=1}^{P_1} \cdots \sum_{p_L=1}^{P_L}\mathbb{B}_{[p_1,\cdots,p_L,q_1,\cdots,q_M]}\mathbb{X}_{[n,p_1,\cdots,p_L]} \\
	&=\sum_{p_1=1}^{P_1} \cdots \sum_{p_L=1}^{P_L}\sum_{s_1=1}^{S_1}\left((\mathbb{D}_{(-)})_{[p_1,\cdots,p_L,s_1,q_2,\cdots,q_M]} \right. \\
	& \times \left. \mathbf{V}_{(1)s_1q_1}\mathbb{X}_{[n,p_1,\cdots,p_L]}\right)   \\
	&=\sum_{s_1=1}^{S_1} \sum_{p_1=1}^{P_1} \cdots \sum_{p_L=1}^{P_L}
	\\ &\left((\mathbb{D}_{(-)})_{[p_1,\cdots,p_L,s_1,q_2,\cdots,q_M]}\mathbb{X}_{[n,p_1,\cdots,p_L]} \right)\mathbf{V}_{(1)s_1q_1}.
\end{split}
\end{equation}
By equation \eqref{eq:post_v2}, we have
\begin{equation}
\label{eq:post_v3}
\hat{\mathbb{Y}}=\mathbb{D} \times_{L+1} \mathbf{V}_{(1)}.
\end{equation}
Combine \eqref{eq:tucker_xn} and equation \eqref{eq:post_v3} resulting in 
\begin{equation*}
\mathbb{{Y}}_{(2)}=\mathbf{V}_{(1)}\times (\mathbb{D}_{(\mathcal{R}\times \Ccal)})^T+\mathbb{E}_{(2)},
\end{equation*}
where $\mathbb{{Y}}_{(2)}\in \Rbb^{Q_1\times N\prod_{m=2}^{M}Q_m}$ is the matricization of tensor $\mathbb{Y}$ and $\mathbb{D}_{(\mathcal{R}\times \Ccal)} \in \Rbb^{N\prod_{m=2}^{M}Q_m \times S_1}$ is the matricization of tensor $\mathbb{D}$.	
\subsection*{Conditional posterior distributions given training fraction $b$}
\label{sec:post_givenb}
Given the dimension $\bm{\theta}= (R_1, \cdots, R_L,S_1, \cdots, S_M)$ of the core tensor $\mathbb{G}$, and the training fraction $b$, we first derive 
the full conditional posterior distributions  of $\{\mathbf{U}_{(l)}\}_{l=1}^{L}$, $\{\mathbf{V}_{(m)}\}_{m = 1}^{M}$, $\mathbb{G}$, $\sigma^2$ in closed forms.  Without loss of generality, we first derive the full conditional posterior distribution of $\mathbf{U}_{(1)}$.  The full conditional posterior distributions of $\{\mathbf{U}_{(2)},\cdots,\mathbf{U}_{(L)}\}$ can be derived in the same manner. 

By equation \eqref{eq:post_u1}, we have
\begin{equation*}
vec\mathbb{Y}= \mathbb{C}_{(\mathcal{R}\times \Ccal)} \times \ vec \mathbf{U}_{(1)}+vec\mathbb{E} \,,
\end{equation*}
where $\mathcal{R}=\{N,Q_1,\cdots,Q_M\}$, and $\mathcal{C}=\{R_1,P_1\}$.
And combined with the idea that $\mathbf{U}_{(1)}$ is from a  distribution proportional to $p(\mathbb{Y} \mid \{\mathbf{U}_{(l)}\}_{l=1}^{L}, \{\mathbf{V}_{(m)}\}_{m=1}^{M}, \mathbb{G},\sigma^2,\bm{\theta} )^{b}\times p(\mathbf{U}_{(1)} \mid \bm{\theta}) $, we have,  the posterior distribution of $vec\mathbf{U}_{(1)}$ given all other parameters and $b$ is normal distribution. That is
\begin{equation}
\label{eq:post_u1_b}
vec\mathbf{U}_{(1)} \ \sim \ N(\bm{\mu}'_U, \mathbf{\Sigma}'_U)
\end{equation} where,
\begin{align*}
\mathbf{\Sigma}'_U & = \left(\frac{b\times \mathbb{C}_{(\mathcal{R}\times \Ccal)}^T\mathbb{C}_{(\mathcal{R}\times \Ccal)}}{\sigma^2}+\mathbf{\Sigma}_U^{-1} \right)^{-1} \,,\\
\bm{\mu}'_U &= \mathbf{\Sigma}'_U \left(\frac{b\times \mathbb{C}_{(\mathcal{R}\times \Ccal)}^Tvec\mathbf{Y}}{\sigma^2}+\mathbf{\Sigma}_U^{-1}\bm{\mu}_U\right) \,.
\end{align*}

We then derive the conditional posterior distributions of 
$\mathbf{V}_{(m)}$ given $\sigma^2$, $\{\mathbf{U}_{(l)}\}_{l=1}^{L}$, $\mathbf{V}_{(k)}$ for $k\neq m$, and $\mathbb{G}$. Without loss of generality, we derive the full conditional posterior distribution of $\mathbf{V}_{(1)}$ below. 

Denote the contracted product of the tensor  $\mathbb{G}\times_1 \mathbf{U}_{(1)} \cdots \times_L \mathbf{U}_{(L)} \times_{L+2} \mathbf{V}_{(2)} \cdots \times_{L+M} \mathbf{V}_{(M)}$ and tensor $\mathbb{X}$ by a new tensor $\mathbb{D}$, where $\mathbb{D}\in \Rbb^{N\times S_1\times Q_2\times \cdots \times Q_M}$. We then matricize $\mathbb{D}$ into a matrix $\mathbb{D}_{(\mathcal{R}\times \Ccal)} \in \Rbb^{N\prod_{m=2}^{M}Q_m \times S_1}$ and write 
\begin{equation}
\label{eq:post_v11}
\mathbb{{Y}}_{(2)}=\mathbf{V}_{(1)}\times (\mathbb{D}_{(\mathcal{R}\times \Ccal)})^T+\mathbb{E}_{(2)},
\end{equation}
where $\mathbb{{Y}}_{(2)}\in \Rbb^{Q_1\times N\prod_{m=2}^{M}Q_m}$ is the matricization of tensor $\mathbb{Y}$.
Let $\mathbf{\tilde{Y}}=\mathbb{{Y}}^T_{(2)}$, given $\mathbf{V}_{(1)}$ follows a normal distribution with a diagonal covariance matrix, 
we can rewrite \eqref{eq:post_v11} as 
\begin{equation*}
vec\tilde{\mathbf{Y}} = \left(\mathbf{I}_{Q_1}\otimes\mathbb{D}_{(\mathcal{R}\times \Ccal)}\right) \times vec\mathbf{V}_{(1)}^T + vec\left(\mathbb{E}_{(2)}\right)^T \,,
\end{equation*}
where $\mathbf{I}_{Q_1}$ denotes an identity matrix of size $Q_1$. Given the prior distribution of $vec\mathbf{V}_{(1)}$ is a normal $N(\bm{\mu}_{V},\bm{\Sigma}_V)$ with diagonal $\bm{\Sigma}_V$, the prior distribution of $vec\mathbf{V}_{(1)}^{T}$ is also a normal distribution $N(\tilde{\bm{\mu}}_V, \tilde{\bm{\Sigma}}_V)$ with a diagonal covariance matrix.
Then the full conditional posterior distribution of $vec{\mathbf{V}_{(1)}}^T$ is also normally distributed:
\begin{equation}
\label{eq:post_v1_b}
p(vec\mathbf{V}_{(1)}^T \mid vec\mathbb{Y},\mathbb{X},\sigma^2, \mathbf{U}_{(l)}, \mathbf{V}_{(m)} \ m\neq 1)\sim \ N(\tilde{\bm{\mu}}_U^{'}, \tilde{\mathbf{\Sigma}}_U^{'}),
\end{equation}
where
\begin{equation}
\begin{split}
	\tilde{\mathbf{\Sigma}}_{V}^{'}&=\left(\frac{b\times\left(\mathbf{I}_{Q_1}\otimes\mathbb{D}_{(\mathcal{R}\times \Ccal)}\right)^T\left(\mathbf{I}_{Q_1}\otimes\mathbb{D}_{(\mathcal{R}\times \Ccal)}\right)}{\sigma^2}+\tilde{\mathbf{\Sigma}}_V^{-1}\right)^{-1} \\
	\tilde{\bm{\mu}}_{V}^{'}&=\tilde{\mathbf{\Sigma}}_{V}^{'} \left(\frac{b\times\left(\mathbf{I}_{Q_1}\otimes\mathbb{D}_{(\mathcal{R}\times \Ccal)}\right)^T vec\mathbf{\tilde{Y}}}{\sigma^2}+\tilde{\mathbf{\Sigma}}_V^{-1}\tilde{\bm{\mu}}_V\right).
\end{split}
\end{equation}

By analogous procedures, we have 
the posterior distribution of $vec \mathbb{G}_{(\mathcal{R}\times \mathcal{C})}$ is also normal distribution with 
\begin{equation}
\label{eq:post_g_b}
\begin{split}
	&\bm{\mu}'_{G} =\mathbf{\Sigma}'_G \left((\mathbf{I}_S\otimes (\mathbb{X}_{(1)}\mathbf{U}))^T(\mathbf{\Sigma_{\tilde{Y}}})^{-1} (b\times vec\mathbf{\tilde{Y}})+\mathbf{\Sigma}_G^{-1}\bm{\mu}_G\right) \\
	& \mathbf{\Sigma}'_G =\\ 
	& \left(b\times(\mathbf{I}_S\otimes (\mathbb{X}_{(1)}\mathbf{U}))^T  (\mathbf{\Sigma_{\tilde{Y}}})^{-1}(\mathbf{I}_S\otimes (\mathbb{X}_{(1)}\mathbf{U}))+(\mathbf{\Sigma}_G)^{-1}\right)^{-1} ,
\end{split}
\end{equation}

where $\mathbf{I}_S$ denotes an $S\times S$ identity matrix with $S=\prod_{m=1}^{M}S_m$. 
And for $\sigma^2$, the posterior given all other parameters and $b$ also follows inverse gamma distribution. That is 
\begin{equation}
\label{eq:post_sigma_b}
\sigma^2 \ \sim \ IG(\alpha', \beta')
\end{equation}
with $\alpha'=\alpha+\frac{b\times NQ}{2}$,  $\beta'=\beta+\frac{b\times\norm{\mathbb{Y}-\langle \mathbb{X,B}\rangle_L}^2_F}{2}$, and $Q=\prod_{m=1}^{M}Q_m$.

\subsection*{Proof for equation \eqref{eq:MH_1}}
\label{sec:Frac_Bayes}
The Fractional Bayes Factor, Eq(11) of \cite{o1995fractional}, is given by
\begin{equation*}
\begin{split}
	B_b(\mathbb{Y}) &= \underbrace{\frac{\int p(\tilde{\bm{\xi}} \mid \tilde{\bm{\theta}}) p(\mathbb{Y} \mid \tilde{\bm{\xi}}, \tilde{\bm{\theta}}) d\tilde{\bm{\xi}}}{\int p(\tilde{\bm{\xi}} \mid \tilde{\bm{\theta}}) p(\mathbb{Y} \mid \tilde{\bm{\xi}}, \tilde{\bm{\theta}})^b d\tilde{\bm{\xi}}}}_{(**)} \\ 
	&\times \frac{\int p(\dbtilde{\bm{\xi}} \mid \bm{\theta}^{(t-1)}) p(\mathbb{Y} \mid \dbtilde{\bm{\xi}}, \bm{\theta}^{(t-1)})^b d\dbtilde{\bm{\xi}}}{\int p(\dbtilde{\bm{\xi}} \mid \bm{\theta}^{(t-1)}) p(\mathbb{Y} \mid \dbtilde{\bm{\xi}}, \bm{\theta}^{(t-1)}) d\dbtilde{\bm{\xi}}} \,.
\end{split}
\end{equation*}
We note that 
\begin{equation*}
\text{(**)} = \frac{\int \overbrace{p(\tilde{\bm{\xi}} \mid \tilde{\bm{\theta}}) p(\mathbb{Y} \mid \tilde{\bm{\xi}}, \tilde{\bm{\theta}})^b}^{(***)} p(\mathbb{Y} \mid \tilde{\bm{\xi}}, \tilde{\bm{\theta}})^{(1-b)} d\tilde{\bm{\xi}}}{\int p(\tilde{\bm{\xi}} \mid \tilde{\bm{\theta}}) p(\mathbb{Y} \mid \tilde{\bm{\xi}}, \tilde{\bm{\theta}})^b d\tilde{\bm{\xi}}} \,.
\end{equation*}
By similar techniques in \cite{o1995fractional}, we can rewrite $p(\mathbb{Y} \mid \tilde{\bm{\xi}}, \tilde{\bm{\theta}})^b$ as $p(\mathbb{Y}' \mid \tilde{\bm{\xi}}, \tilde{\bm{\theta}})$ that is the likelihood based on $\mathbb{Y}'$ which is the training proportion $b$ of data $\mathbb{Y}$. And rewrite (***) as 
\begin{equation*}
\begin{split}
	\text{(***)} &= p(\tilde{\bm{\xi}} \mid \tilde{\bm{\theta}}) p(\mathbb{Y}' \mid \tilde{\bm{\xi}}, \tilde{\bm{\theta}}) \\
	& = p(\tilde{\bm{\xi}} \mid \mathbb{Y}', \tilde{\bm{\theta}}) \int p(\tilde{\bm{\xi}} \mid \tilde{\bm{\theta}}) p(\mathbb{Y}' \mid \tilde{\bm{\xi}}, \tilde{\bm{\theta}})d\tilde{\bm{\xi}} \\
	& =p(\tilde{\bm{\xi}} \mid \mathbb{Y}', \tilde{\bm{\theta}}) \int p(\tilde{\bm{\xi}} \mid \tilde{\bm{\theta}}) p(\mathbb{Y} \mid \tilde{\bm{\xi}}, \tilde{\bm{\theta}})^b d\tilde{\bm{\xi}} \,.
\end{split}
\end{equation*}
Then $B_b(\mathbb{Y})$ becomes
\begin{equation}
\label{eq:MH_1sup}
B_b(\mathbb{Y}) = \frac{\int p(\tilde{\bm{\xi}} \mid \mathbb{Y}', \tilde{\bm{\theta}})p(\mathbb{Y} \mid \tilde{\bm{\xi}}, \tilde{\bm{\theta}})^{(1-b)} d\tilde{\bm{\xi}}}{\int p(\dbtilde{\bm{\xi}} \mid \mathbb{Y}', \bm{\theta}^{(t-1)})p(\mathbb{Y} \mid \dbtilde{\bm{\xi}}, \bm{\theta}^{(t-1)})^{(1-b)} d\dbtilde{\bm{\xi}}} \,.
\end{equation}
And (*) in \eqref{eq:MH_1} is evaluating the Fractional Bayes Factor $B_b(\mathbb{Y})$ at one sample of $\tilde{\bm{\xi}}$ and $\dbtilde{\bm{\xi}}$ instead of integrating out as in \eqref{eq:MH_1sup}.
\subsection*{Full simulation study results of CP method with different ranks}

\setcounter{table}{0}
\renewcommand\thetable{\Alph{section}.\arabic{table}}

\begin{table*}[!htpb]
\centering
\caption{Mean RPE with SD for the CP method with different $R$ values.}
\resizebox{\textwidth}{!}{\begin{tabular}{c|cccccc}
		\hline 
		\textbf{RPE} (SD) & $R = 1$ & $R = 2$ & $R = 3$ & $R = 4$ & $R = 5$ & $R = 6$ \\ 
		\hline 
		\textbf{Uncorrelated Data} & & & & & & \\
		\hline 
		$\theta^*= (4,4,2,2)$, SNR=2 & 0.771(0.102) & 0.627(0.087) &0.529(0.095) & 0.463(0.061) & 0.409(0.036) & 0.377(0.014)  \\ 
		\hline 
		$\theta^*= (3,3,3,3)$, SNR=2 & 0.780(0.081) & 0.623(0.061) & 0.535(0.048) & 0.466(0.038) & 0.422(0.024) & 0.391(0.016)  \\ 
		\hline 
		$\theta^*= (4,4,2,2)$, SNR=5 & 0.718(0.137) & 0.530(0.109) & 0.392(0.072)  & 0.308(0.052) & 0.241(0.039) & 0.209(0.046) \\ 
		\hline 
		$\theta^*= (3,3,3,3)$, SNR=5 & 0.718(0.095) & 0.527(0.074) & 0.414(0.060) & 0.320(0.044) & 0.266(0.032) & 0.222(0.016) \\ 
		\hline 
		\hline 
		\textbf{Correlated Data} & & & & & & \\
		\hline 
		$\theta^*=(4,4,2,2)$, SNR=2 &0.533(0.152)& 0.423(0.095)& 0.382(0.043)& 0.371(0.021)& 0.370(0.011)& 0.377(0.014)  \\ 
		\hline 
		$\theta^*=(3,3,3,3)$, SNR=2 & 0.514(0.120)& 0.429(0.070)& 0.391(0.037)& 0.374(0.019)& 0.370(0.010)& 0.369(0.007)  \\ 
		\hline
		$\theta^*=(4,4,2,2)$, SNR=5 & 0.407(0.170)& 0.266(0.104)& 0.219(0.056)& 0.198(0.034)& 0.190(0.014)& 0.188(0.006) \\ 
		\hline 
		$\theta^*=(3,3,3,3)$, SNR=5 & 0.413(0.187) & 0.280(0.085) & 0.230(0.047) & 0.205(0.026) & 0.194(0.015) & 0.187(0.006) \\ 
		\hline 
\end{tabular}}
\label{tbl:cp1}
\end{table*}

\label{sec:CP_rank}
For each case of the simulation study, we tried the CP method with CP-rank $R$ values from 1 to 6. Note that when $R = 5,6$, the total number of parameters in CP method are $230$ and $276$ respectively which are larger than the true number of parameters (212 for $\bm{\theta}^* = (4,4,2,2)$ cases and 219 for $\bm{\theta}^* = (3,3,3,3$) cases). We calculate the RPE results with different $R$ values for all 50 repeated experiments under all simulation setups. And means and standard deviations of RPEs averaging over the 50 repeated experiments for all 8 simulation setups are shown in  Table \ref{tbl:cp1}.

\bibliographystyle{apalike}
\bibliography{references}
\end{document}